\documentclass[11pt,a4paper]{article}
\usepackage{jheppub}
\usepackage{graphicx}

\def\hbar{\hspace{0pt}\raisebox{1pt}{$-$} \hspace{-7pt} h}

\def\5{\overline 5}

\newcommand{\sun}{\ensuremath{\odot}}
\newcommand{\earth}{\ensuremath{\oplus}}%
\newcommand{\ba}{\begin{eqnarray}}
\newcommand{\ea}{\end{eqnarray}}
\newcommand{\no}{\nonumber}
\newcommand{\be}{\begin{equation}}
\newcommand{\ee}{\end{equation}}
\newcommand{\bea}{\begin{eqnarray}}
\newcommand{\eea}{\end{eqnarray}}

\title{A dark force for baryons}
\date{\today
}
\author{
Michael L. Graesser, Ian M. Shoemaker, and Luca Vecchi}
\affiliation{Theoretical Division T-2, Los Alamos National Laboratory \\ Los Alamos, NM 87545, USA}
\abstract{We suggest the existence of a fundamental connection between baryonic and dark matter. This is motivated by both the stability of these two types of matter as well as the observed similarity of their present-day densities. A unified genesis of baryonic and dark matter is natural in models in which the baryon number is promoted to a spontaneously broken local gauge symmetry. This is illustrated in a specific class of SUSY models using the Affleck-Dine mechanism. The dark matter candidate in these scenarios is charged under the baryon gauge symmetry and must have a mass around the GeV scale to give the correct present-day abundance.  We discuss constraints from B-factories, LEP, mono-jet searches at the Tevatron, and dark matter direct detection experiments.  A baryonic dark force is shown to be consistent with all data for mediators as light as the GeV scale. 

}
\begin{document}
\maketitle
%

\section{Introduction}

The stability of the proton in the standard model (SM) is the consequence of an \emph{accidental} symmetry of the renormalizable Lagrangian. Generic UV completions of the SM are hence expected to induce proton decay. Even if ad hoc symmetries are introduced in the renormalizable formulation of these theories, the question remains: what prevents the proton from decaying through higher dimension operators? 
We interpret the stability of the proton as one of the few experimental facts that physics beyond the SM should explain.

Global symmetries are commonly viewed as accidental at low energies, as they are expected to be violated by quantum gravity effects. In order to guarantee the conservation of a certain global number one should therefore appeal to more fundamental, dynamical arguments.

A simple way to ensure the stability of the proton is to promote the baryon number to a local symmetry.  In this paper we explore the consequences of gauged baryon number, spontaneously broken at the weak scale. Within the framework proposed here, proton stability is ensured at all orders in perturbation theory, and is not spoiled by non-renormalizable effects. 

The idea of gauging the baryon number has been studied in a number of papers in the eighties and nineties~\cite{rajpoot,he,foot,MC1,MC2,Bailey,Carone}, and more recently in~\cite{Wise,Wise2,Dong:2010fw,Ko:2010at,baryonbump,Perez:2011dg,Perez:2011pt,Lebed:2011fw}. The main difference between our work and those already present in the literature is that here the lepton and baryon sectors are treated in a totally \emph{asymmetric} way, as we gauge baryon number to protect the proton stability,  but allow explicit lepton number violation. 

This approach is phenomenological in spirit, since there are no observations suggesting a fundamental relationship between leptons and baryons. For example, the stability of the electron is ensured in any theory where it is the lightest electrically charged fermion. This follows from an interplay between electric charge conservation and kinematics without any need for additional dynamical assumptions. Furthermore, the smallness of the neutrino masses can be elegantly arranged via the see-saw mechanism, where lepton number is explicitly violated.

In contrast, there are indications of a connection between the baryonic sector and dark matter.  Such a connection is suggested by two observational facts: (1) both dark matter and protons are stable on cosmological timescales without any obvious explanation a priori; and (2) precise cosmological observations reveal that their present-day abundances are remarkably similar, being only a factor of about 5 apart.

Interestingly, the promotion of baryon number to a local symmetry requires the existence of a new anomalous global symmetry. We find that under rather general dynamical conditions this new physics provides a natural dark matter candidate. Moreover, the genesis of visible and dark matter in fact \emph{unifies} in such a way that their relic abundance can be similar for comparable masses.  Our scenario overlaps with the ``asymmetric dark matter" scenario \cite{Kaplan,Agashe:2004bm,ADM,hooper,Hylogenesis,Darko,Dodelson,Xo,Falko,adm9,adm12,adm5,AWIMP,Frandsen:2011kt,MarchRussell:2011fi,Davoudiasl:2011fj,Cui:2011qe}, although here the asymmetry is not generated in one sector and then subsequently transferred to the other.  Rather, in our framework both asymmetries are generated simultaneously~\cite{pangenesis,cogenesis}. 

We focus on SUSY realizations, although this is not a necessary ingredient in our treatment. The main advantages offered by SUSY realizations are that (i) the theory is technically natural and calculable up to very high scales;  and (ii) the genesis of matter can be elegantly explained by the Affleck-Dine mechanism \cite{AD,DRT,DRT2}.

The main features of our framework are as follows:
\begin{itemize}

\item We promote baryon number to a gauge symmetry $U(1)_{B}$. Anomaly cancellation implies the existence of new exotic quarks with their own anomalous $U(1)$ global symmetry. 

\item To facilitate the decay of exotic quarks, a SM singlet chiral superfield $X$ is introduced.  The dark matter $X$ has gauge baryon interactions and is stable, since it is the lightest particle charged under the new anomalous $U(1)$. 

\item The DM plus the visible sector have a \emph{single}, linearly realized, nonanomalous accidental global symmetry $U(1)_D$.  Baryogenesis requires a primordial asymmetry in $U(1)_D$, which necessarily implies the existence of an asymmetry in the dark sector.

\item This general setup has supersymmetric flat directions, allowing for the {\it simultaneous} generation of dark and baryonic asymmetries via the Affleck-Dine mechanism.

\item The proper DM abundance is achieved via the annihilation mode to quarks mediated by the new baryonic gauge boson $Z_{B}$. 

\item The $Z_{B}$ gauge boson also mediates interactions between dark matter and nuclei.  Though the coupling to nuclei is vectorial, the coupling to DM is model dependent.

\end{itemize}

The outline of the paper is the following. 

In Section~\ref{models} we present a class of UV complete models in which the baryon number is gauged, and discuss the phenomenological viability of these models.  Models in which the dark matter has either a purely vectorial or purely axial coupling to the $U(1)_B$ gauge boson are presented in Section \ref{UV}.

In Section~\ref{baryogenesis} we see that both the visible and dark sectors have comparable primordial asymmetries, and show that typical perturbative realizations of our scenarios require the DM candidate to have a mass between sub-GeV to tens of GeV. 

In Section~\ref{Bounds} we present a general analysis of the bounds arising from B-factories, LEP and Tevatron, and direct detection experiments. Here we work under the assumption that the only remnant of the dark sector are the baryon gauge field and a DM particle  carrying charged under the gauged baryon number.  If vectorially coupled to the $U(1)_B$ gauge boson, the DM must be around the GeV scale to evade direct detection constraints, whereas axial couplings lead to novel momentum and velocity suppression such that direct detection bounds are much weaker.  Constraints from mono-jet searches at the Tevatron are strong when the mediator is produced on-shell.

\begin{figure}
\begin{center}
\includegraphics[width=2.9in]{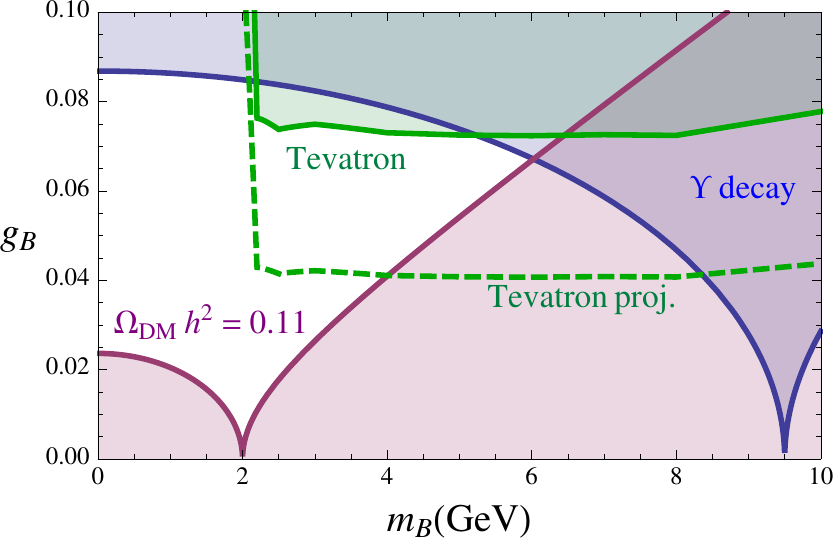}
\includegraphics[width=2.9in]{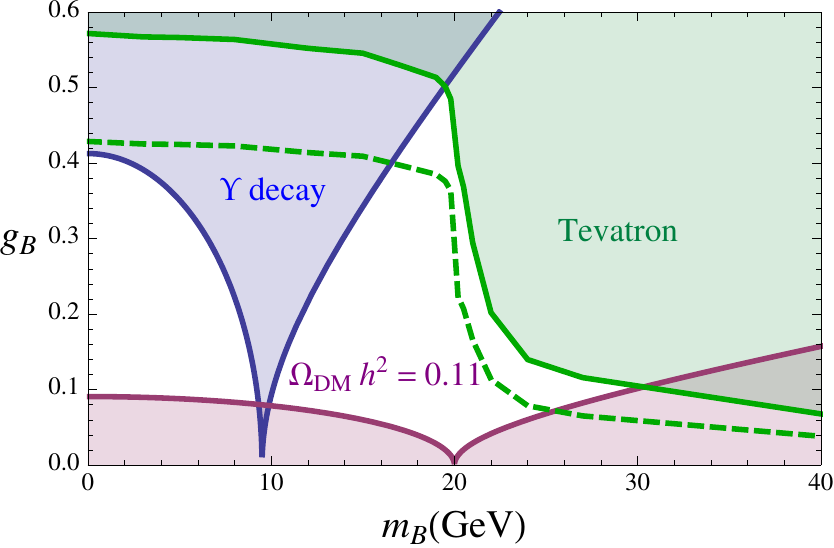}
\caption{\small 
Constraints on a dark matter candidate. {\em Left}: $m_X=1$ GeV and a purely vector coupling with $q_V=4/3$. {\em Right:}
$m_X=10$ GeV and a purely axial coupling with $q_A=4/3$.  
The filled areas are excluded by $\Upsilon \rightarrow invisible$ (left) or $\Upsilon \rightarrow hadrons$ (right)  and the CDF monojet search. Tevatron projections assuming $10$ fb$^{-1}$ of integrated luminosity are also shown (dashed).
The solid red line indicates the values for $(m_B, g_B)$ required to get the right DM abundance for a symmetric DM species (WIMP cross section). Asymmetric DM species must lie above the red curve. 
\label{Tev100-axial}}
\end{center}
\end{figure}

Our main numerical results are summarized in Fig. \ref{Tev100-axial} for two benchmark examples. The first panel shows the allowed region for a 1 GeV dark matter particle with a purely vector coupling to the $U(1)_B$ gauge boson.  The second panel shows the allowed region for a 10 GeV dark matter particle having a purely axial coupling to the $U(1)_B$ gauge boson.

While our cosmological predictions for the ratio of the dark matter asymmetry to the baryon asymmetry are sensitive to the UV completion, our collider and direct detection bounds are independent of the specific realization for the cancellation of the $U(1)_B$ anomalies.


\section{A class of UV complete models}
\label{models}

\subsection{The $q'$-sector and the baryon Higgs sector}

In this section we will present a class of UV complete models.  Clearly these models do not exhaust all viable realizations of our scenario, as many alternatives are possible.

The SM baryon number $U(1)_{B_q}$ is anomalous under the electroweak $SU(2)_W\times U(1)_Y$ gauge group, and in order to promote it to a local symmetry $U(1)_B$ one needs to introduce exotic fermions transforming under a chiral representation of the SM symmetry. A simple solution is to introduce $N$ new generations of quarks and leptons with the new quarks $q'$ carrying gauge baryon number $B(q')=-\frac{1}{N}$~\cite{MC2}. 

We focus on supersymmetric realizations of this program, in which baryon number violation would otherwise arise at the renormalizable level. We thus add to the MSSM  $N$ extra generations of quarks and leptons superfields. Table~\ref{SM'} summarizes the charge assignments of the exotic fields, with $i=1,\dots,N$. One can readily see that the gauge anomalies vanish. The right handed neutrinos $\nu_c'$ have been introduced to make the new fields sufficiently heavy to evade current bounds.

We also include a minimal, vectorial baryon Higgs sector composed of two chiral superfields $S,\overline{S}$. Here $B(S)$, the $U(1)_B$ charge of the baryon Higgs fields $S,\overline{S}$, is arbitrary, and will only be constrained by the requirement of a sufficiently long lifetime for the proton and the DM, see Sec.~\ref{sPDM}. 
\begin{table}
\begin{center}
\begin{tabular}{|c||c|c|c|c|}
\hline
& $SU(3)_C$ & $SU(2)_W $ & $U(1)_Y$ & $U(1)_B$\\ 
\hline\hline
$Q'_i$ & $3$ & $2$ & $+\frac{1}{6}$ & $-\frac{1}{N}$\\  
\hline
${u_c'}_i$ & $\bar 3$ & $1$ & $-\frac{2}{3}$ & $+\frac{1}{N}$\\  
\hline
${d_c'}_i$ & $\bar 3$ & $1$ & $+\frac{1}{3}$ & $+\frac{1}{N}$ \\  
\hline\hline
$L'_i$ & $1$ & $2$ & $-\frac{1}{2}$ & $0$ \\  
\hline
${\nu_c'}_i$ & $1$ & $1$ & $0$ & $0$\\
\hline
${e_c'}_i$ & $1$ & $1$ & $+1$ & $0$\\  
\hline
\hline
$S$ & $1$ & $1$ & $0$ & $+B(S)$\\  
\hline
$\overline{S}$ & $1$ & $1$ & $0$ & $-B(S)$\\  
\hline
\end{tabular}
\caption{Example of field content of the $q'$-sector and $U(1)_B$ Higgs sector. Here the index $i$ runs from 1 to $N$, where $N$ is the number of new generations. The scalars $S,\overline{S}$ are responsible for breaking the gauge baryon symmetry. \label{SM'}}
\end{center}
\end{table}

In the absence of $R$-parity the usual lepton number violating operators $W_{\not L}=H_uL+LLe_c + QLd_{c}+\nu^\prime _c\nu^\prime_c$, where summation over the SM and new generations of leptons and quarks is implied, are present. While the couplings of these operators cannot be too large in order to evade the bounds from lepton flavor violations, one expects that Majorana masses for the SM neutrinos can be generated quite naturally. As long as any of the couplings of these operators is switched on, the LSP will be unstable. For example, the neutralino and the gravitino will both decay into leptons, e.g. $\chi_0\rightarrow e^+\tilde e^-\rightarrow e^+e^-\nu$ and $\tilde G\rightarrow\gamma\nu$. These processes might take place outside the detector, implying a collider phenomenology similar to the  MSSM.

The most general renormalizable superpotential for the system (Table~\ref{SM'}) is a sum of the $q$- and $q'$-sector superpotentials, the lepton number violating $W_{\not L}$, plus the baryon Higgs sector term $W_S=m_SS\overline{S}+S\overline{S}\nu_c$, and formally reads
\ba
W_{q}+W_{q'}+W_{\not L}+W_S.
\ea

The phenomenology of the new fermion generations share some features with those of a fourth SM generation. However there are some differences. 

As with the fourth generation, here new quarks with masses $M_{q'}$ in the few hundred GeV scale are expected to be consistent with the EW precision measurements provided a splitting of the weak doublets is present and $N$ is not large~\cite{kribs}\cite{Holdom}. 

With a 4th generation 
 the coupling of the Higgs boson to gluons is enhanced due to a triangle loop involving these quarks. Recent Higgs boson searches from the LHC appear to exclude the possibility of a 4th generation \cite{Gunion,ATLASHiggs2011}.  However, as noted in \cite{Gunion}, the Higgs boson may simply be heavier than 200 GeV. This is possible in extensions to the minimal supersymmetric Standard Model, such as through the addition of singlets \cite{Barbieri}~\footnote{More generally, the enhancement of the Higgs production is not generic in these models for the following reasons. 
First, 
in order to cancel the $U(1)_B$ anomalies, the 4th generation only has to be charged under {\em some} $SU(3)$. This $SU(3)$ does not have to be the same as $SU(3)_C$.  
Secondly, another possibility is that the 4th generation may be vector-like with respect to the Standard Model interactions, but chiral with respect to the $U(1)_B$ in order to cancel the $U(1)_B$ anomalies 
\cite{Perez:2011pt}. In this case the 4th generation gets a mass from both $U(1)_B$ symmetry breaking and electroweak symmetry breaking. Then the coupling of the Higgs boson to gluons depends on the ratio of the part of the quark masses arising from electroweak symmetry breaking to the total quark masses, and a suppression may be possible.}. 

Another crucial difference between our $q'$ sector and a typical 4th generation is that the gauge symmetry $U(1)_B$ forbids mixing between the new quarks and the SM quarks.  This implies the decay modes of these exotic quarks, discussed in the following Section,  are {\em not} like a standard 4th generation quark. 
The 4th generation leptons however, can have mass mixing with the lighter generations, so they can decay through the weak interactions, or through the renormalizable lepton-number violating $R_p$-violating interactions that we assume are present.

\subsection{The dark matter}
\label{2.2}

Given the gauge baryon charge assignments and field content of the Table~\ref{SM'}, the $q'$-quark generations will not mix with $q$. This implies that the new generations will carry their own baryon number $U(1)_{B_{q^\prime}}$, and will hence be stable unless new interactions are introduced. 

A simple way to evade cosmological problems associated with new stable, charged particles, is to introduce a (vectorial) sector of particles $X$ coupling the MSSM and the new generations of quarks that makes the reactions ``$q'$-sector$\rightarrow q$-sector$+X$" possible. 
By construction the $X$ sector will also carry the $U(1)_{B_{q^\prime}}$ number and can include a natural dark matter candidate. 

In general the coupling of the DM to the baryonic gauge boson is chiral, 
\ba
D^\mu X=\partial^\mu X+ig_B \left( q_{V} + q_{A} \gamma^{5}\right)Z_{B}^\mu X.
\ea
In Appendix~\ref{UV} vector and chiral models are presented, in which the coupling of the DM after $U(1)_B$ symmetry breaking is either purely vector or axial.  We will now proceed with two illustrative vector models defined at the effective level. 

First consider the case in which the dark matter chiral superfields $X^{\pm}$ have the following representation under $SU(3)_{C}\times SU(2)_{W}\times U(1)_{Y} \times U(1)_{B}$:
\ba
X^\pm\sim\left(1,1,0,\pm\left(\frac{1}{3}+\frac{1}{N}\right)\right)~~~~~~~~~{\rm Model ~I} 
\ea
Then the lowest order terms in the superpotential following from the assumed field content is $W_{tot}=W_{q}+W_{q'}+W_{\not L}+W_S+W_X^{\rm{eff}}$, with the last term given by
\ba\label{nonrenI}
W_X^{\rm{eff}}&=&m_XX^+X^-+\frac{H_uQ'u_cX^+}{\Lambda}+\frac{H_dQ'd_cX^+}{\Lambda}+\frac{H_uQu_c'X^-}{\Lambda}+\frac{H_dQd_c'X^-}{\Lambda}\\\no
&+&\dots
\ea
The higher dimension operators will serve as ``transfer operators" - mixing the primordial dark and visible asymmetries.   
They also allow the processes $q'\rightarrow qX$, 
provided the channels are kinematically open.  Flavor changing processes are safe provided $v/ \Lambda \lesssim 10^{-2}$.

Next consider the case in which the DM superfields have the representation:
\ba
X^\pm\sim\left(1,1,0,\pm\left(\frac{2}{3}-\frac{1}{N}\right)\right)~~~~~~~~~{\rm Model ~II} 
\ea
Now the leading interactions with the SM are described by 
\ba\label{nonren}
W_X^{\rm{eff}}&=&m_XX^{-}X^{+}+\frac{u_cd_cd_c'X^{+}}{\Lambda}+\dots.
\ea
As before, the higher dimension operator will be an effective transfer operator and also allow for 
 $q^\prime \rightarrow qqX$. 
Moreover, although the transfer operator $u_cd_cd_c'X^{+}$ violates flavor, it does not contribute to meson oscillations at 1 loop. The dominant contribution is in fact to purely hadronic heavy meson decays. These are expected to be under control for $\Lambda$ larger than a few TeV.

A common feature of both of these models is the unusual phenomenology of the lightest exotic quark. The heavier exotic quark decays to the lightest $q^\prime$ through charged current weak interactions, and therefore has the same experimental signatures as of a conventional 4th generation quark. But since the 4th generation of quarks has no mass mixing with the SM quarks, the lightest $q^\prime$ decays through the transfer operator, and therefore only to quarks and missing energy. The experimental signatures for the lightest $q'$ are therefore significantly different from standard fourth generation fermions. Here the lightest $q'$ always produce missing energy and never a $W$ or any charged leptons.

\subsection{Stability of the proton and the dark matter}
\label{sPDM}

Before gauging $U(1)_B$, the theories defined in the previous subsections -- and specifically $W_{tot}$ -- have 3 accidental global $U(1)$'s:
the $U(1)_{B_q}$ number under which the ordinary quarks and the DM are charged; the $U(1)_{B_{q^\prime}}$ number under which the new quarks and the DM are charged; and a nonanomalous symmetry $U(1)_{B_S}$ carried by the singlets $S,\overline{S}$ under which $S$ has unit charge. These charge assignments are summarized in Table \ref{table:globalcharges}.

\begin{table}
\begin{center}
\begin{tabular}{|c|c|c|c|}
\hline
 & $B_q$  & $B_{q^\prime}$ & $B$  \\ \hline 
Model I & & & \\ \hline  
$q$ &$+\frac{1}{3}$ & 0  & $+\frac{1}{3}$ \\ \hline 
$q^\prime$ & 0&  $- \frac{1}{N}$ & $ - \frac{1}{N}$ \\ \hline 
 $X$ &   $+\frac{1}{3}$ & $\frac{1}{N}$ &  $+\frac{1}{3}+ \frac{1}{N}$ \\ \hline
\hline
 & $B_q$  & $B_{q^\prime}$ & $B$  \\ \hline 
Model II & & & \\ \hline  
$q$ &$+\frac{1}{3}$ & 0  & $+\frac{1}{3}$ \\ \hline 
$q^\prime$ & 0&  $- \frac{1}{N}$ & $ - \frac{1}{N}$ \\ \hline 
 $X$ &   $+\frac{2}{3}$ & $-\frac{1}{N}$ &  $+\frac{2}{3}- \frac{1}{N}$ \\ \hline
\end{tabular}
\caption{Accidental  global and $U(1)_B$ charge assignments of the quarks, exotic quarks and dark matter in Models I and II. $S$ is neutral under the accidental global symmetries $B_q$ and $B_{q^\prime}$. \label{table:globalcharges}}
\end{center}
\end{table}

The gauge symmetry $U(1)_B$ is a linear combination of these $U(1)$'s:
\ba\label{gaugeB}
B=B_q+B_{q'}+B(S)B_S.
\ea
The vacuum of $S,\overline{S}$ violates $U(1)_B$ but leaves the global, anomalous symmetries $U(1)_{B_q}$ and $U(1)_{B_{q'}}$ unbroken. This ensures that, at the renormalizable level, the stability of the proton is guaranteed even after the spontaneous breaking of the baryon gauge number. Similarly, the lightest state carrying the number $U(1)_{B_{q'}}$, here $X$, will also be stable at the renormalizable level.

Let us now discuss the stability of both proton and DM beyond the renormalizable action, and more generally in any theory with gauged $U(1)_B$. The following is a slight generalization of the analysis presented in~\cite{MC2} and applies to theory of gauged baryon number.

We assume that the lightest low energy states in the model are the proton and mesons of QCD, the leptons, the photon, and the DM. The most general process mediating proton decay must involve a single proton in the initial state and a number of mesons, leptons and photons, and DM fields in the final state, and it can be effectively described by a higher dimensional operator of the form
\ba\label{PD}
\Psi L^\alpha X^\beta S^\gamma \mathcal{O},
\ea
where $\Psi\sim qqq$ is the proton interpolating field, $L$, $X$, and $S$ the lepton, DM, and baryon Higgs operators, respectively, whereas $\alpha,\beta,\gamma$ are (positive or negative) integers. The operator $\mathcal{O}$ contains arbitrary powers and derivatives of the $U(1)_B$ singlets $(\overline qq)$, $(\overline LL)$, $(\overline XX)$, $(\overline{S} S)$, and terms involving the photon. Note that the power $\alpha$ is not determined by electric neutrality, since the meson fields as well as the lepton bilinears in the operator ${\cal O}$ might carry a nonzero charge. The numbers $\alpha, \beta$ might instead be constrained by Lorentz invariance. For example, if the DM is fermionic we should require $\alpha+\beta=odd$ while if the DM is bosonic $\alpha=odd$. However, these constrains are not relevant to our analysis.

The crucial constraint in fact comes from the requirement that the operator~(\ref{PD}) be compatible with the baryon gauge symmetry. This in turn implies  
\ba\label{decayP}
1+ \beta B(X)+ \gamma B(S)=0,~~~~~~~~~~~~~~~~{\rm{(implies~proton~decay)}}
\ea
with $B(S)$ and $B(X)$ the $U(1)_B$ gauge charges of the $S$ field and the DM respectively. If no integers $\beta,\gamma=0,\pm1,\pm2,\dots$ exist such that the relation~(\ref{decayP}) is satisfied, perturbative proton stability in the model is ensured even beyond the renormalizable level. Nonperturbative effects might involve nonlocal operators, and are entirely negligible at the scales of interest.

Stability of the DM can be proved beyond the renormalizable level in a similar way, by examining operators of the form 
\be
X \Psi^{\delta} S^{\epsilon} L^{\nu} \mathcal{O}'.  
\ee
DM stability is guaranteed in a model in which no integers $\delta,\epsilon=0,\pm1,\pm2,\dots$ exist such that the constraint
\ba\label{decayDM}
B(X)+\delta+\epsilon B(S)=0~~~~~~~~~~~~~~~~{\rm{(implies~DM~decay)}}
\ea
is satisfied.

In general, these requirements can be satisfied for generic $U(1)_B$ charges.  We now describe sufficient conditions for stability.  When $m_{X} > m_{p}$ ($\beta =0$) proton stability is ensured for integer values of $B(S)$ with absolute value greater than unity, whereas DM stability is guaranteed for fractional $B(X)$.  The conditions on the charges are different when $m_{p} > m_{X}$ ($\delta =0$).  Here DM stability is ensured for integer values of $B(S)/B(X)$ with absolute value grater than unity. Proton stability is then guaranteed with the additional requirement of integer $B(X)$ with absolute value greater than unity.

\section{A unified genesis of dark matter and baryons}
\label{baryogenesis}

A direct consequence of having the Higgs sector be vectorial is the existence of an accidental nonanomalous $U(1)_D$ with the generator 
\be
\label{D}
D = B_{q} + B_{q'}. 
\ee
 This symmetry 
is the \emph{only} global symmetry of these model, and a primordial asymmetry $\eta_D$~\footnote{Throughout the paper the asymmetries $\eta_s=(Y_s-Y_{\bar s})$ are conventionally defined as the difference between the yield of particles $Y_s$ and antiparticles $Y_{\bar s}$ of the species $s$.} in this quantity is required by baryogenesis. In such a framework, comparable primordial asymmetries for $U(1)_{B_q}$ and the number $U(1)_{B_{q'}}$ will be simultaneously generated, as we will now show. A common genesis of DM and baryons can therefore be viewed as a natural consequence of this construction. Similar considerations have 
recently been considered in ~\cite{pangenesis,cogenesis}.

Before continuing, it is worth noting that in models \emph{without} low-scale lepton number violation one anticipates the existence of more global symmetries involving baryon, lepton, and dark numbers, and hence more primordial asymmetries. In that case, the asymmetries of the dark and baryonic matter are not trivially related, as in general both the present-day DM and the ordinary baryon numbers receive contributions from a primordial lepton asymmetry. In the models considered here there is no such ambiguity: the baryon asymmetry is not related to a primordial lepton asymmetry, but rather the baryon and dark matter asymmetries are related to a primordial asymmetry in the non-anomalous $U(1)_D$ symmetry.

\subsection{Asymmetry generation via Affleck-Dine }

An especially  simple possibility for the simultaneous generation of the dark and visible asymmetries in the early universe is Affleck-Dine baryogenesis~\cite{AD,DRT,DRT2}. 
Recall that in supersymmetry,  there are typically  directions $\phi$ in scalar field space along which the potential is identically zero in the supersymmetric limit.  
 If the flat direction acquires a negative mass squared during inflation from SUSY breaking, then it will have
a large vacuum expectation value in the early universe due to the  balance between this soft mass and higher dimensional terms either in the superpotential and/or SUSY breaking $A-$ terms in the scalar potential.  
Since some of these ``flat directions'' will carry a non-zero charge under any global symmetries, such as $D$, a non-zero and large asymmetry is generically generated once the Affleck-Dine (AD) condensate $\phi$ begins to oscillate freely about its vacuum minimum (typically when $H \sim m_{soft}$), provided the $A-$terms in the scalar potential  violate the global symmetry and have a $CP$ violating phase that is not identical to the one in the inflaton.  Below $H \sim m_{soft}$ the higher dimension operators that break the global symmetry become irrelevant, and the resulting primordial asymmetry in $D$ remains conserved. 
  In this way a  large asymmetry in a conserved global number  can be generated~\cite{DRT,DRT2}.  
  
Eventually the AD condensate will evaporate due to collisions with the plasma generated from reheating \cite{AD} and transfer the asymmetry to baryons. For this to occur several factors must be considered.  First, the  evaporation process should conserve the global number, otherwise the asymmetry gets washed out.
Next, the 
condensate should not evaporate before a large asymmetry has built-up. Too rapid of an evaporation can happen from the scattering of the AD condensate off of the dilute plasma generated when the inflaton first begins to decay. This leads to a constraint that depends on the dimension of the operator that violates the global symmetry, the reheat temperature $T_R$ and the mass scale $M$ suppressing the higher dimension operator.  For $n>4$ this constraint is easily satisfied, while $n=4$ marginally allowed \cite{DRT,DRT2}.~\footnote{$n$ is defined below.}
 
The last  
 consideration is whether the AD field has evaporated before the era of electroweak symmetry breaking.  
While one typically expects the decay of the AD condensate long before the era of electroweak symmetry breaking~\cite{DK}, the same is not true for the $Q$-balls~\cite{Coleman} that may form from the fragmentation of the condensate~\cite{KS}.  Although in gauge-mediated models large $Q$-balls are stable, and thus a dark matter candidate in their own right~\cite{firstQ,KS}, here we assume for simplicity that they are unstable.  However the fact that their decay proceeds only through the surface, leads to long lifetimes~\cite{evap} that may decay after EW symmetry breaking~\cite{EM,GravQ}.  Thus we will also examine the possibility that the asymmetry is deposited to the plasma below the EW scale.

Next we show that both models I and II introduced in Sec.~\ref{2.2} have flat directions that violate the $U(1)_D$. 
To that end, note that from the very definition of $U(1)_B$ and $U(1)_D$ - see~(\ref{gaugeB}) and~(\ref{D}) - it follows that all possible gauge invariant operators will be automatically $U(1)_D$ invariant unless nontrivial powers of the Higgs field $S$ are present. In other words, any flat direction $\phi$ relevant for the Affleck-Dine mechanism must involve nontrivial powers of $S$ and therefore must violate $U(1)_{B_S}$ and $U(1)_D$. 
We will assume these operators which involve $S$ are suppressed by a scale $M$ large compared to the scale $\Lambda$ appearing in models I and II.  

Since up to this point the $U(1)_B$ charge of $S$ are  arbitrary, the  
 spirit here is to assign it a gauge charge so the candidate flat direction is $U(1)_B$ gauge invariant.  
In doing this one also has to check the flat direction is not lifted by the superpotential interactions (\ref{UVI}) for model I, or (\ref{UVII}) for model II. 

Once a gauge charge for $S$ has been chosen, one can then check whether the conditions (\ref{decayP}), (\ref{decayDM}) for proton and dark matter decay are ever satisfied. As was shown in Sec.~\ref{sPDM} the conditions for absolute stability of both particles depends on the relative hierarchy of the proton and DM masses.  In what follows, we focus on the case $m_{X} > m_{p}$.

There are a number of possible classical flat directions; we will parameterize the holomorphic and gauge-invariant flat direction\footnote{We follow the notation of \cite{DRT,DRT2}.} as $\phi^n$. We assume SUSY breaking generates CP violating and $U(1)_D$ breaking $A$ terms $\propto \phi^n/M^{n-3}$ in the potential, and also generates a negative mass squared during inflation. We also assume the superpotential has a term 
$W \propto \phi^n/M^{n-3}$ which stabilizes the flat direction.

\underline{Model I}: 
One can have
the following $n=5$ flat direction, 
 $\phi^5=u^{\prime} _c d^{\prime}_c d _c X S$ (where we assign $B(S)=-3$). 

 \underline{Model II}:
In this model the analogous flat direction is 
 $\phi^5=u^{\prime} _c d^\prime_c d _c{\overline{X}}S$, but where we assign $B(S)=-2$.
 
In both of these models the dark matter and the proton are stable since the charges satisfy the conditions in Sec.~\ref{sPDM} for $m_{X} > m_{p}$.  In the second example neutron oscillations can occur via the dimension-10 operator $\Psi\Psi\langle S\rangle$.

\subsection{Primordial asymmetries}

As the universe cools, the $U(1)_D$-violating operators (suppressed by $M$) decouple around $H \sim m_{soft}$ and the asymmetry $\eta_D$ freezes in. Below this scale the theory has a non-anomalous $U(1)_D$ symmetry.

Eventually the AD condensate evaporates and the asymmetry $\eta_D$ is transferred to the plasma.
We define the temperature at which this occurs to be 
 $T_\phi$. Whether this occurs above or below the scale $T^*=O(100$ GeV$)$ at which the EW sphalerons shut-off is model-dependent. 
 By definition below $T^*$ the sphalerons are inefficient, yet  $U(1)_D$ preserving operators of the form $q^\alpha{q'}^\beta X^\gamma{\cal O}'''$ $-$ such as occurs in (\ref{nonrenI}) and (\ref{nonren}) $-$ are generally still in chemical equilibrium. Then eventually the exotic quarks $q^\prime$ freeze-out and decay. 
When they freeze-out their abundance is Boltzmann suppressed, which implies 
that when they decay 
they transfer very little asymmetry to the DM particles.

If the operators $q^\alpha{q'}^\beta X^\gamma{\cal O}'''$
also preserve the $U(1)_{B_q}$ and $U(1)_{B_{q'}}$ global symmetries, which is precisely realized in Models I and II, then below $T^*$ these charges become separately good quantum numbers. In this case the determination of the asymmetries simplifies.
Since this scenario is realized in models I and II we will proceed under this assumption, though it isn't necessary.  At the end of this Section we will comment on how the analysis is different if this assumption is invalidated.
 
Continuing under the assumption that $U(1)_{B_q}$ and $U(1)_{B_{q'}}$ are global symmetries 
 for $T<T^*$, then 
  at present times the transfer is complete and 
\ba 
\label{present-Bq}
\eta_{B_{q^\prime}}=B_{q\prime}(X)\eta_X
\ea 
Similarly, the $U(1)_{B_{q}}$ asymmetry at late times reads 
\ba 
\label{present-Bqp} 
\eta_{B_{q}}=\eta_{B_{vis}}+B_{q}(X)\eta_X 
\ea 
with $\eta_{B_{vis}}$ referring to the pure SM contribution to the baryon asymmetry. We thus see that the ratio between the SM baryon asymmetry and the DM asymmetry today is finally
\ba\label{note1}
\frac{\eta_{B_{vis}}}{\eta_X}=\frac{\eta_{B_{q}}}{\eta_{B_{q^\prime}}}B_{q^\prime}(X)-B_{q}(X).
\ea
There is another way to understand why the second term is present. Since the dark matter is charged under the generalized baryon asymmetry $B_q$, its contribution must be subtracted out to obtain the amount of baryon charge left in the conventional baryons.   

The crucial step in determining the present-day abundances is the determination of $\eta_{B_{q}}/\eta_{B_{q'}}$.  Let us distinguish between two possible scenarios.

If the asymmetry $\eta_D$ is communicated to the thermal bath at a temperature $T_\phi<T^*$ the present-day asymmetries $\eta_{B_{q,q'}}$ are entirely determined by the charge assignments of the flat direction/D-ball. For example, given a flat direction with $U(1)_{B_q}$ charge $B_q(\phi)$ and an $U(1)_{B_{q'}}$ charge $B_{q'}(\phi)$ one finds
\ba\label{1}
\frac{\eta_{B_q}}{\eta_{B_{q'}}}=\frac{B_q(\phi)}{B_{q'}(\phi)}~~~~~~~~~\rm{for}~T_\phi<T^*.
\ea
As an example, the flat direction $\phi^5={u^\prime} _c {d^\prime} _c d_c \overline{X}S$ introduced earlier for model II has, for $N=1$, $\eta_{B_q}/\eta_{B_{q'}}=-\frac{1}{3}$. Assuming that the transfer operator in~(\ref{nonren}) is still active at these temperatures, and plugging this result in~(\ref{note1}) we find $\eta_{B_{vis}}/\eta_X=-\frac{1}{3}$.

If instead $T_\phi>T^*$, the two asymmetries $\eta_{B_q}$ and $\eta_{B_{q'}}$ vary as a function of the temperature until the scale $T^*$, while the asymmetry $\eta_D$ remains conserved. The present-day baryon asymmetries in this case are determined by equilibrium thermodynamics at the critical temperature $T^*$, and the relation~(\ref{1}) will typically not be correct.  Moreover the precise relation will also depend on a monotonically decreasing function $f(m_{i}/T)$ which accounts for the Boltzmann suppression in the density of a species of mass $m_{i}$. 

\begin{figure}
\begin{center}
\includegraphics[width=4.0in]{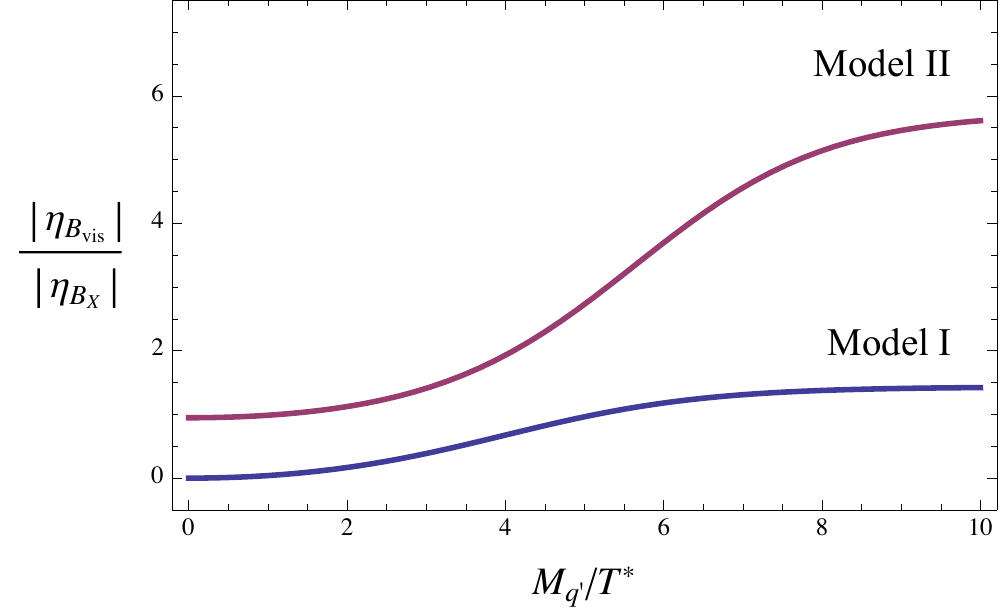}
\caption{\small The ratio~(\ref{note3f'}) (model I) and  (\ref{note2f'}) (model II)  between the primordial asymmetries of ordinary baryons and DM for $N=1$ as a function of a comman mass $M_{q^\prime}/T^*$ for the $q^\prime$-sector fields. \label{fIandII}}
\end{center}
\end{figure}

In general the present-day abundance depends on whether the sphalerons decouple above or below the electroweak symmetry breaking phase transition $T_{EW}$.  In Appendix \ref{chempot} we perform a chemical potential analysis and show that for $T_\phi>T^*$ and $T^*<T_{EW}$ the visible to baryonic asymmetry ratios are 
\ba\label{note3f'}
\frac{\eta_{B_{vis}}}{\eta_X}&=&-\frac{9 (-1 + f_{q^\prime}) N (21 + 5 f_{q^\prime}N))}{ 99 + (33 + 618 f_{q^\prime}) N + f_{q^\prime} (17 + 135 f_{q^\prime}) N^2)}
~~\hbox{(model I)}
\ea
\ba\label{note2f'}
\frac{\eta_{B_{vis}}}{\eta_X}&=&-\frac{9 N(21+f_{q^\prime}(45+2(3+5f_{q^\prime})N)}{99+N(-66+f_{q^\prime}(618+(-7+135f_{q^\prime})N))}  ~~~~\hbox{(model II)} 
\ea
where we have abbreviated $f_{q^\prime} \equiv f(M_{q^\prime}/T^{*})$ and set $f_X=1$. These results are shown in Figure \ref{fIandII} for $N=1$ and $0<M_{q'}/T^*<10$.

If on the other hand the sphalerons decouple above the EW phase transition then we show in Appendix~\ref{chempot} that one just sets $f_{q^\prime} =1$ in (\ref{note3f'}) and (\ref{note2f'}).  

The most striking feature in Figure \ref{fIandII} is the sensitivity of the present-day ratio to $f_{q'}$. In particular, the present-day ratio for Model I vanishes for $f_{q^\prime}=1$, which can occur if the sphalerons decouple above the electroweak phase transition, or simply if $M_{q^\prime} \ll T^*$.  Unfortunately the 
sphaleron scale $T^*$ is not known with precision. Common lore says that $m_W\lesssim T^*\lesssim O($few$)\times100$ GeV. Meanwhile, perturbativity in the Yukawa sector forces us to estimate a zero-temperature upper bound of order $M_{q^\prime}\lesssim300-400$ GeV. A conservative measure of how large $M_{q^\prime}/T^*$ can be is thus $M_{q^\prime}/T^*=\frac{400~{\rm{GeV}}}{m_W}\approx5$, though in Figure \ref{fIandII}  the plots are extended beyond this estimate.

In models where there is no  $U(1)_{B_q}$ and $U(1)_{B_{q'}}$ accidental symmetries below $M$, and only $U(1)_D$, then the analysis proceeds in two steps. First one solves for $\eta_D$ at the temperature $T=T^*$. Below that scale the sphalerons are inefficient by definition, yet baryon number violating operators connecting the quarks, exotic quarks and dark matter may still be in thermal equilibrium down to lower temperatures. If so, one re-does the analysis  down to the temperature at which these interactions decouple and then evaluates the asymmetries, using the results of the analysis at $T=T^*$ as a boundary condition.

\subsection{A light dark matter candidate}
\label{lightDM}

Below a model-dependent temperature the quantities $\eta_{B_{q^\prime}}$ and $\eta_{B_q}$ are separately conserved. However, the relative contributions of the chiral fermions and the DM changes with time. At some characteristic scale $T\lesssim M_{q^\prime}$ the heavy quarks will freeze-out and then decay via the operators~(\ref{nonrenI}) and (\ref{nonren}). After freeze-out their contribution to the $U(1)_{B_{q^\prime}}$ asymmetry is exponentially suppressed, leaving the totality of $\eta_{B_{q^\prime}}$ to the DM particle. As we saw in the previous Section, then at late times the $U(1)_{B_{q^\prime}}$ asymmetry is just given by 
\ba\label{lateq'}
\eta_{B_{q^\prime}} & \rightarrow & B_{q^\prime}(X)\eta_X~~~~~~~~~~~~~({\rm{at~late~times}}). 
\ea
Similarly,  
\ba\label{lateq}
\eta_{B_{q}} & \rightarrow & \eta_{B_{vis}}+B_{q}(X)\eta_X~~~~~~({\rm{at~late~times}}),
\ea
with $\eta_{B_{vis}}$ being the primordial asymmetry carried by the ordinary SM quarks. 

The present-day baryon and DM abundances are finally explained if
\ba
\frac{m_X}{m_n}\left(\frac{Y_{+} +Y_{-}}{Y_{+} -Y_{-}}\right)= \left | \frac{\eta_{B_{vis}}} {\eta_X}\right| \,\frac{\Omega_X}{\Omega_B},
\ea
with $m_n$ the nucleon mass, $\eta_{B_{vis}}/\eta_X$ given by eq.~(\ref{note1}), $Y_{\pm}\equiv n^{\pm}/s$ being the present-day number densities of DM particles and antiparticles normalized to the entropy density $s$, while $\Omega_B, \Omega_X$ are the baryonic and DM density in the universe. The ratio of antiparticles to particles depends exponentially on the product $\eta_X\langle\sigma_{ann}v\rangle$, such that $Y_{-} \ll Y_{+}$ when the annihilation cross section is slightly above the value required for a symmetric species~\cite{GS,AWIMP,Drees2011}.

For either vector or axial couplings, the dominant annihilation mode for a fermionic DM $X$ is through an s-channel $Z_B$ exchange into SM quarks. Hence we find that the annihilation cross section times the relative particle velocity at freeze-out is:
\ba\label{ann}
\langle\sigma_{ann}v\rangle&=&\sum_f\frac{N_c}{2\pi}m_X^2\left(\frac{g_B^2}{3m_B^2}\right)^2\frac{\left(2+\frac{m_f^2}{m_X^2}\right) \left( q_{V}^{2} + \frac{2T}{m_{X}} q_{A}^{2}\right) }{\left(1-\frac{4m_X^2}{m_B^2}\right)^2+\frac{\Gamma_B^2}{m_B^2}}\,\,\sqrt{1-\frac{m_f^2}{m_X^2}}. 
\ea
where the sum extends over all kinematically allowed 2-SM quarks states. 
Note that the vector and axial couplings give rise to s- and p-wave annihilation respectively. We do not consider the two gauge boson final state that is open when $m_X>m_B$.

Without loss of generality we can assume $0 \le Y^{-} \le Y^{+}$, which yields
\ba\label{mX<}
m_X\leq \left | \frac{\eta_{B_{vis}}} {\eta_X}\right|\,\frac{\Omega_X}{\Omega_B}\,m_n.
\ea
Whether the flat direction evaporates above or below the sphaleron scale, the DM mass in this class of models is required to be at the GeV scale.  For example, assuming that $T_\phi>T^*$,   model I (\ref{nonrenI}) requires $m_X \lesssim 10$ GeV, while model II~(\ref{nonren}) requires $m_X\lesssim30$ GeV, both for $N=1$, as can be inferred from Fig. \ref{fIandII}. If instead $T_\phi<T^*$ our prototype flat direction $\phi^5=u_c'd_c'd_c{\overline{X}}S$ implies $m_X\lesssim2$ GeV.  The masses of $X$ and $Z_{B}$ can be naturally related to the weak scale in a theory in which the DM is chiral under $U(1)_{B}$.  In this case the DM will generally have both axial and vector couplings to the $Z_B$.

We will see in Section~\ref{DD} that the direct detection signals are very large when the dominant interactions for the DM are mediated by a vectorially coupled $U(1)_B$ force. If the annihilation mode is dominated by the $Z_B$ exchange, then the only allowed region of the parameter space is for DM masses below $1-2$ GeV for nonzero vectorial couplings.  Purely axial couplings of dark matter to the $Z_{B}$ lead to suppressed scattering rates such that direct detection experiments are not constraining.

\subsection{Direct detection}
\label{DD}

In the following we make the minimal assumption that the physics relevant for annihilation and direct detection are both mediated by the baryonic gauge boson $Z_B$. This assumption leads to a lower bound on the elastic scattering cross section of nuclei relevant for direct detection experiments. Of course if the $Z_B$ exchange \emph{does not} dominate the annihilation process the bounds become much less restrictive.  

The event rate of DM-nucleus scattering is
\be \frac{ dR}{d E_{R}} = \frac{N_{T} \rho_{\sun}}{m_{X}} \int_{|\vec{v}| > v_{min}} d^{3} v ~v f(\vec{v},\vec{v}_{\earth}) \frac{d \sigma}{d E_{R}}, \ee
where $v_{min} = \sqrt{m_{N}E_{R}}/\sqrt{2} \mu_{N}$, $m_N$ is the target nucleus mass,  $N_{T}$ is the number of target nuclei in the detector, $\rho_{\sun}$ is the local DM density, 
$v_{\earth} \simeq 220$km/s is the velocity of the earth,
and $\mu_{N}$ is the DM-nucleus reduced mass.  We assume that the DM in the galaxy can be taken as an equilibrated system such that the velocity distribution has a one-to-one relationship with the density distribution. Thus following~\cite{DT2} we take
\be
f_{k}(v) \propto \left[\exp \left( \frac{v_{esc}^{2}-v^{2}}{kv_{0}^{2}}\right)-1\right]^{k}, \ee
where $1.5 < k< 3.5$ is a shape parameter related to the outer density profile. 

We first consider the case when only the vector coupling is present $(q_A=0)$. Here the matrix element is velocity independent at leading order and the differential cross section is
\be \frac{d \sigma}{d E_{R}} = \frac{m_{N} A^{2}}{2 \pi v^{2}} \left(\frac{q_{V} g_B^2}{m_B^2}\right)^2 F^{2}(E_{R}), \ee
where $F(E_{R}) = 3 j_{1}(q r_{0})/(qr_{0}) e^{-(qs)^{2} {\rm fm}^{2}/2}$ is the Helm form factor with $r_{0}^{2} =(1.44 A^{2/3}-5 )~{\rm fm^{2}}$.

In the case of nonzero vector couplings the bounds from direct detection are significant and generally require low DM mass.  As a representative example we consider a $1~$GeV DM mass with a single nucleon scattering cross section
\ba \label{dddm}
\sigma_{Xn}&=&\frac{\mu_{n}^2}{\pi}\left(q_V\frac{g_B^2}{m_B^2}\right)^2,\\\no
&\approx& 4 \times 10^{-37}{\rm cm}^{2} \left(\frac{g_{B}}{0.045}\right)^4 \left(\frac{4~\rm{GeV}}{m_{B}}\right)^{4} \left(\frac{\mu_{n}}{0.5}\right)^{2} q_{V}^{2}
\ea
where $\mu_{n}=m_Xm_n/(m_X+m_n)$ is the reduced mass of the nucleon/DM system.

The light DM regime is interesting in that direct detection experiments have yet to significantly constrain it.  Since direct detection experiments detect recoil events above an energy threshold $E_{thr}$, there is a corresponding target-dependent velocity threshold $v_{thr} = \sqrt{ m_{N} E_{thr}/2\mu_n^{2}}$. Whenever $v_{thr}$ exceeds the maximum annual velocity $v_{\earth} + v_{esc}$, no recoil events are possible and, correspondingly, no meaningful bound on the DM elastic scattering cross section can be inferred. 

 The strongest bounds on light DM (sub O(3 GeV))  are from the CRESST-I experiment, which had a very low energy threshold $E_{thr} = 0.6$ keV and a relatively light target. In the second line of Eq.(\ref{dddm}) we have assumed values consistent with the CRESST-I data~\cite{cresst} for a 1 GeV DM candidate.
 For DM masses less than about 1 GeV, one finds that 
  with CRESST-I parameters  there exists no bound, since the required threshold velocity exceeds the escape velocity. If we include astrophysical uncertainties related to possible deviations from a Maxwellian distribution in the high DM velocity tail relevant for these cases~\cite{DT1}\cite{DT2}, we can fairly say that for DM masses in the region $m_X\lesssim1-2$ GeV there is no relevant bound from direct detection experiments. This point is especially relevant for the class of models we propose, since as we showed in Section~\ref{baryogenesis}, in order to obtain the correct present-day abundance the DM mass must be at or below the GeV scale.

For heavier DM and vector couplings the resulting nucleon-DM cross section is typically very large.  For example, for a weak scale DM mass the relation~(\ref{dddm}) is compatible with the current direct detection bounds only if annihilation occurs nearly on-resonance ($m_X\sim m_B/2$); see for example~\cite{Wise}. 

If on the other hand only the axial coupling is present $(q_V=0)$, the spin-independent rate has a velocity and momentum suppression,
\be \label{axdd} \frac{d \sigma}{d E_{R}} = \frac{m_{N}A^{2}}{8 \pi v^{2}}\left(\frac{q_{A} g_B^2}{m_B^2}\right)^2 \left[ 4v^{2} - \frac{q^{2}}{m_{N}^{2}m_{X}^{2}} \left(m_{N}^{2} + 2m_{N}m_{X} - m_{X}^{2}\right)\right] F^{2}(E_{R}),
\ee
where the momentum transfer is $q^{2} = 2m_{N} E_{R}$.
Then the direct detection bounds are currently not constraining. 
This is because the axial coupling contribution to the rate is suppressed by the dark matter velocity and target recoil momentum. 
The velocity and momentum dependence in the axial case lead to a suppression in the overall rate compared to the vector case, varying from around $10^{-6}$ for light DM to $10^{-5}$ for TeV scale DM, with the weakest suppression $(10^{-5})$ occurring for $k=1.5$.

We determined the total rate for the axial scenario, obtained by integrating  (\ref{axdd}) over all velocities and  recoil energies, for 
DM masses between 1 GeV and 1 TeV, and following \cite{DT2}, for  $k$ in the range 1.5 to 3.5.   While we did not perform an exhaustive scan of the parameter space, for the points we checked the rate 
 was never above the bounds quoted by XENON100 \cite{Aprile:2011hi}, CDMS \cite{cdms} and CRESST-I \cite{cresst}.  
For example, currently the strongest experimental bound is $\sigma < 2 \times 10^{-45}$ cm$^2$ from XENON100 \cite{Aprile:2011hi} and occurs at $m_{DM} \sim 50$ GeV. With 
parameters $(g_B,m_B)=(0.025,150 \hbox{ GeV})$ and $m_X=50$ GeV, chosen such that the correct present-day abundance of DM was obtained, we find for $k=1.5$
the total rate to be a factor
of 2 below the XENON100 bound.  On the other hand, for $(g_B,m_B)=(0.1,30 \hbox{ GeV})$ and 
$m_X=10$ GeV, the rate is at least a factor of 20 below the bound from XENON100.  The predicted rate is lower for larger $k$ and larger DM masses.

\section{Baryonic dark forces and collider experiments}
\label{Bounds}

We now turn to the collider phenomenology of our models. 

We assume that the lightest non-SM fields are the dark matter $X$ and the baryonic gauge boson $Z_{B}$. Under these assumptions the field $X$ has a mass $m_X$ parametrically lighter than the mass scale $M_{q'}$ characterizing the $q'$-sector  and our framework can be effectively described in a model-independent way by the following low energy Lagrangian:
\ba\label{eff}
{\cal L}&=&{\cal L}_{SM}+\overline X\gamma^\mu D_\mu X-m_X\overline XX\\\no
&-&\frac{1}{4}\left(Z_B^{\mu\nu}Z_B^{\mu\nu}- 2c_Z s_w Z_B^{\mu\nu}Z^{\mu\nu}+2c_\gamma c_w Z_B^{\mu\nu}A^{\mu\nu}\right)+\frac{m_B^2}{2}Z_B^\mu Z_B^\mu+\dots
\ea
where the dots refer to nonrenormalizable operators suppressed by powers of $M_{q'}$ or $\Lambda$, while $Z_B^{\mu\nu}$, $Z^{\mu\nu}$, and $A^{\mu\nu}$ are the field strength of the baryon gauge field $Z_B$, the $Z^0$ boson, and the photon, respectively. The Lagrangian ${\cal L}_{SM}$ refers to the standard model (SM) in which the baryon number has been gauged. The SM quarks have a vectorial coupling to the $Z_B$ boson with charge $\frac{1}{3}$ whereas the field $X$ is assumed to carry both a vectorial and axial charge:
\ba
D^\mu X=\partial^\mu X+ig_B \left( q_{V} + q_{A} \gamma^{5}\right)Z_{B}^\mu X.
\ea

For $m_B > 2 m_X$ the width of the new gauge boson is approximately
\ba
\label{width}
\Gamma_B=\left[\frac{N_fN_c}{9}+q_V^2\left(1+ 2 \frac{m^2_X}{m^2_B}\right) \lambda^{1/2}+ q_{A}^{2} \lambda^{3/2}\right]\frac{g_B^2}{12\pi}m_B,
\ea
where $\lambda = 1-4 m^2_X/m^2_B$,  $N_c=3$ is the number of colors and $N_f$ is the number of SM flavors lighter than $m_B/2$ (which we assumed to be massless for simplicity), and $g_B$ is the gauge coupling.

Because the new force is leptophobic, the mixing coefficients $c_\gamma,c_Z$ will play a major role in the discussion of the LEP bounds on this scenario. In the models considered here no large log corrections to these quantities are generated since Tr$(BY)=0$ -- with $B$ and $Y$ the baryon and hypercharge charges while Tr sums over all representations -- and one finds $c_\gamma=c_Z=0$ above $M_{q'}$~\cite{MC2}.  Then the mixing coefficients $c_{\gamma,Z}$ are only generated by SM quark loops below the scale $M_{q^\prime}$ and one obtains \cite{MC2}
\ba
\mu\frac{dc_{\gamma}}{d\mu}&=&-\frac{ g_Be}{18 \pi^2 c_w} \left[2N_u- N_d\right],\\\no
\mu\frac{dc_{Z}}{d\mu}&=&-\frac{g_B e}{32 \pi^2s^2_wc_w} \left[3 (N_d-N_u)+ 4s_w^2\left(2N_u-N_d\right)\right],
\ea
with $c_{\gamma,Z}(M_{q^\prime})=0$. Here $N_{u,d}$ is the number of up/down quark flavors active below the scale $\mu$. For example, $N_u=N_d=3$ for $M_{q'}>\mu>m_t$ and $N_u=2$, $N_d=3$ for $m_b<\mu<m_t$. Hence, for $M_{q^\prime}=300$ GeV we find
\ba
\label{mixc}
c_Z(m_Z)&\approx&+2 \times 10^{-2}\,g_B, \nonumber\\
c_{\gamma}(m_{\Upsilon}) &\approx& +9 \times 10^{-3}\, g_B. 
\ea

In the following we will analyze the experimental constraints on the above scenario~(\ref{eff}). We will assume that the running of the coupling $g_B$ is negligible, which is a very good approximation. In the class of models we presented the coupling $g_B$ in fact stays perturbative up to remarkably high scales.

\subsection{Collider constraints}

Collider experiments severely constrain beyond the standard model physics. In particular, very stringent bounds apply to new force carriers coupled to the light leptons. Far less stringent bounds apply to leptophobic forces (for early references see~\cite{Langacker}). 

As we showed in Section~\ref{baryogenesis}, the DM in the class of models we propose is typically light, with a mass $m_X=O(1$ GeV$)$ to $O(10$'s GeV$)$. We will therefore be interested in this regime when discussing the bounds on the model~(\ref{eff}). 

Previous studies of the baryonic force mediated by $Z_B$ were presented in~\cite{MC1,MC2, Bailey, Carone}, and more recently in ~\cite{quevedo}. 
We follow \cite{MC1,MC2} and \cite{Carone} and update the experimental measurements where available.

A dramatic consequence of gauging baryon number is the introduction of  a dijet resonance in $p p/\overline{p} \rightarrow Z_B \rightarrow jets$.  
Remarkably, for $m_B \lesssim O(100$ GeV$)$ this signature poorly constrains the model, with  \cite{MC1, MC2} finding  
$g_B = O(1)$ allowed by Tevatron analyses.  Older UA2 data is slightly more constraining in this mass range~\cite{ua2}.  Using MadGraph/MadEvent~\cite{MG} we simulated $p \overline{p} \rightarrow Z_{B}\rightarrow jj$, with a branching ratio of 1, at $\sqrt{s} =540$ GeV in the mass range $130 ~{\rm GeV} \le m_{B} \le 160 ~{\rm GeV}$.  We find that at leading order the cross section is below the UA2 bound provided $g_{B} \lesssim 0.5$.  We expect the bound to be weaker for masses below 130 GeV since the background rises faster than the signal. 

The new ingredient present in our analysis compared to these previous references \cite{MC1,MC2, Bailey, Carone,quevedo}  is the existence of a coupling between the baryonic force and the DM, typically signaled by ``missing energy". This fact turns out to be crucial for the B-factory and Tevatron bounds, but not relevant for LEP physics.  Because of this new interaction the constraints become much stronger than those for a model in which the $Z_B$ only couples to quarks. 

Tevatron constraints on dark matter are obtained from searches for events having a high $p_T$ jet and missing energy ~\cite{Fermi,Irvine,Irvine2, Davoudiasl:2011fj}. Compared to \cite{Fermi,Irvine,Irvine2}, here we analyse the implications of these null searches in the context of a specific model. This allows us to be precise about the dependence of the width on the underyling model parameters, and in particular the gauge coupling, which turns out to be important when the mediator can be produced on-shell.

The results of the current and previous sections are summarized in 
Fig.~\ref{Tev100-axial}. Here all the relevant constraints are shown for a DM fermion of mass $m_X=1$ GeV, for the case of a vector coupling, and of mass $m_X=10$ GeV for the case of a purely axial coupling. 

We see that for the purely vector coupling case the parameter space consistent with the present-day DM abundance lies in the regime $g_B<0.1$ and small mediator mass regime $m_B\lesssim6$ GeV. This conclusion would only be slightly affected by an improvement of the Tevatron mono-jet bounds, with the result of pushing the mediator to even smaller masses $m_B\lesssim4$ GeV. In general for a 1 GeV DM mass the strongest constraints come from collider searches.

In the case of a purely axial coupling and $m_X > 5$ GeV the bounds from the invisible width of the Upsilon do not apply, but modifications to the hadronic width of the Upsilon are relevant (and independent of the dark matter properties).
We see that compared to the vector case, here 
the parameter space consistent with the present-day DM abundance is larger, with both a larger $g_B$ coupling and mediator mass  $m_B\lesssim30$ GeV allowed.
An improvement of the Tevatron mono-jet bound will probably require the gauge boson mass to be $m_B \lesssim 2 m_X$.

We now turn to discussing these collider constraints in more detail. 

\subsubsection{B-factories}

Constraints from 
B-factories fall into two signature classes, depending on whether or not the dark matter is kinematically accessible.

In the first, the dark matter is lighter than $\sim $ 5 GeV. 
Here these experiments are
especially suited to detect light dark matter production~\cite{Chang,McElrath,Yeghiyan}.  Remarkably strong bounds for the theory~(\ref{eff}) follow from the physics of the $\Upsilon(1S)$ meson. Its invisible width has been severely bounded by a recent measurement of the BaBar collaboration~\cite{BaBar1},
\ba
{\cal BR}(\Upsilon(1S)\rightarrow\textrm{``invisible"})<3\times10^{-4}
\ea
at the $90\%$ confidence level. For $m_B$ not too close to $m_\Upsilon\simeq9.5$ GeV and $m_X<m_\Upsilon/2\simeq4.7$ GeV we therefore require 
\ba\label{Uinv}
\frac{{\cal BR}(\Upsilon(1S)\rightarrow\textrm{``invisible"})}{{\cal BR}(\Upsilon(1S)\rightarrow \mu^+\mu^-)}=\left[\frac{g_B^2}{e^2}\frac{m_\Upsilon^2}{m_B^2-m_\Upsilon^2}\right]^2\left(q_{V}^2+ q_{A}^{2}\right)<1.2\times10^{-2}.
\ea
In estimating the branching ratio we have made the conservative approximation of ignoring the dark matter mass.  This constraint is shown in Fig. \ref{fig2}.

The next set of constraints in this class comes from bounds on $e^+e^-
\rightarrow\gamma\mu^+\mu^-$~\cite{CLEO1, BaBar2} and
$e^+e^-\rightarrow\gamma$+``nothing" \cite{CLEO2,CLEO3,BaBar3,BaBar4},
provided by the BaBar and CLEO collaborations. 
The former process bounds
the $Z_{B}$-photon mixing to be $c^2_w c_{\gamma}^{2}(m_{\Upsilon}) \lesssim
10^{-5}$~\cite{essig1,essig2} when the $Z_{B}$ can be produced on-shell and decays to muons with essentially $100\%$ branching fraction. 
In our scenario the rate is further suppressed by  $ {\cal BR}(Z_B \rightarrow \mu^+ \mu^-) \propto c^2_w c_{\gamma}^{2}$, simply because the dominant decay of the $Z_B$ is to quarks (or dark matter if allowed).  The constraint is then $c^4_w c_{\gamma}^{4}(m_{\Upsilon}) \lesssim 10^{-5}$, or using (\ref{mixc}), $g_B \lesssim O(1)$ which is weaker than other constraints. If the $Z_B$ is off-shell then the constraint becomes even weaker. 

The rate for $e^+e^-\rightarrow\gamma$+``nothing" from dark matter production is easily related to $e^+e^- \rightarrow\gamma\mu^+\mu^-$. 
Since the experimental bounds on
these two processes are comparable, we expect the constraint from $e^+e^-
\rightarrow\gamma +$ ``nothing"  to be  $c^2_w c_{\gamma}^{2}(m_{\Upsilon}) {\cal BR}(Z_B \rightarrow X \overline{X}) \lesssim
10^{-5}$ when the $Z_B$ is produced on-shell. 
Setting the branching ratio to one, 
this constraint translates into  a bound of $g_B \lesssim 0.4$.  When $m_{B} \gtrsim 10$
GeV  the $Z_B$ is off-shell and  this bound is weakened by a factor $\sim
m_{B}/\Gamma_{B}$. Using (\ref{mixc}) and (\ref{width}) we find that this
constraint implies $g_B \lesssim O(1)$.  In both cases these two bounds are
always weaker than the bounds arising from the invisible and hadronic widths of the
upsilon~(\ref{Uinv}), as can be seen by inspecting Fig.~\ref{fig2}.  

If the dark matter is not kinematically accessible then other constraints become important.  

The strongest constraint when the dark matter is heavier than $\sim 5$ GeV is obtained from  the contribution of the $Z_{B}$ to the hadronic
width of the $\Upsilon$~\cite{MC1,MC2}
\be \Delta R_{\Upsilon} = \frac{4}{3} \left[ \frac{g_{B}^{2}}{e^{2}}
\frac{m_{\Upsilon}^{2}}{m_{B}^{2} - m_{\Upsilon}^{2}}
+\left(\frac{g_{B}^{2}}{e^{2}} \frac{m_{\Upsilon}^{2}}{m_{B}^{2} -
m_{\Upsilon}^{2}} \right)^{2} \right],
\ee
where  $R_{\Upsilon} \equiv \Gamma(\Upsilon \rightarrow
{\rm hadrons})/\Gamma(\Upsilon \rightarrow \mu^{+}\mu^{-})$. Here the dominant
constraint comes from corrections to the decay into two
jets~\cite{Carone}. The bound  ${\cal BR} (\Upsilon \rightarrow jj) <
0.053$ at $95\%$ CL~\cite{argus}, is shown for reference in
Fig.~\ref{fig2}.

Additional constraints arise from lighter quarkonia, but they turn out to be relevant only very close to threshold. We will ignore these bounds in the following (for an analysis of these constraints see~\cite{Carone}), and always assume that $m_B$ is not close to any of the known C-odd mesons.

The $Z_B$ boson is also expected to alter many of the processes used to extract measurements of the strong coupling. In updating the analysis of~\cite{MC1,MC2,Bailey,Carone} we verified that these corrections are negligible for a coupling $g_B$ of weak strength.

\subsubsection{LEP bounds}

The DM particle $X$ and the gauge boson $Z_B$ couple to leptons only through the loop-induced mixing introduced in~(\ref{eff}). The LEP experiment is thus expected to constrain the magnitude of $c_Z$ and $c_\gamma$, and in turn of $g_B$. The constraints obtained here are independent of the properties of the dark matter. 

\begin{figure}
\begin{center}
\includegraphics[width=4.0in]{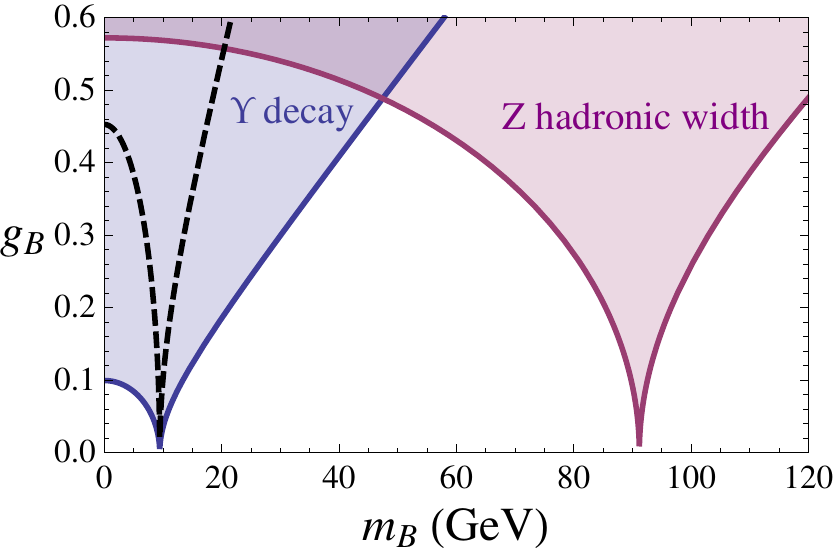}
\caption{\small The white area represents the allowed region in the parameter space $(m_B,g_B)$. These areas are excluded by the experimental bounds on the invisible and hadronic (dashed) widths of the $\Upsilon(1S)$ meson and the hadronic width of the $Z^0$ boson. The bounds from the invisible $\Upsilon$ width have been plotted for $q_{V,A}^2=1$, while the bound from the hadronic $Z^{0}$ width is independent of $q_{V,A}$.  The coupling is bounded by $g_B\lesssim0.5$ in the low mass regime $m_B<m_Z$. The region in the figure near $m_B \sim m_Z$ is not accurate since here the mixing between $Z_B$ and $Z$ is large, which is beyond the validity of our small mixing approximation.  \label{fig2}}
\end{center}
\end{figure}

In agreement with \cite{MC1,MC2}, we find the measured $Z^0$ hadronic width provides the dominant constraint. Focusing on the correction $Z^0\rightarrow Z_B\rightarrow \overline{q}q$ we have for $| c_Z(m_Z)| \ll 1$ and  
$|m_B - m_Z | \gg \Gamma_{Z,B}$
\ba\label{had}
\frac{\Delta\Gamma_{had}}{\Gamma_{had}}\simeq 1.193\,\frac{g_B}{\sqrt{4 \pi}}\,c_Z(m_Z)\frac{m_Z^2}{m_Z^2-m_B^2}+O(c^2)<\pm1.1\times10^{-3}.
\ea
There are additional corrections $O(g_B^2)$ from loop contributions to the quark-$Z^0$ vertex and from the on-shell production $Z^0\rightarrow q\bar qZ_B$ (for $m_B<m_Z$), see~\cite{MC2}. In agreement with~\cite{MC2} we found that the assumption that~(\ref{had}) constitutes the dominant correction is a conservative approximation. Taking $c_Z(m_Z)=0.02 g_B$ from eq. (\ref{mixc}) and assuming a 2-sigma deviation we plot the bound~(\ref{had}) on the plane $(m_B,g_B)$ in Figure~\ref{fig2}. 

We have also looked at the induced correction to the $Z^0$ mass, to the forward-backward asymmetry, and to the measurement of $\alpha_s(M_Z)$ extracted from $Z^0$ decay, and find the derived constraints to be weaker. The exotic contributions to the invisible $Z^0$ width are proportional to $c_Z^2$ and are also very small.

\subsubsection{Mono-jet searches at the Tevatron}

Hadron collider experiments also provide stringent bounds on the model~(\ref{eff}). The most distinctive signals involve missing energy in the final state plus a single jet, while di-jet processes $p\bar p\rightarrow Z_B^*\rightarrow jj$ are expected to be overwhelmed by the QCD background for light $Z_B$. We will thus focus on the following process
\ba\label{proc}
p\bar p\rightarrow X\overline{X}j.
\ea

The CDF collaboration has searched for new physics contributions to the process~(\ref{proc}). In~\cite{CDF} the collaboration reported the detection of $8449$ events with an integrated luminosity of  $1$ fb$^{-1}$, after the cuts discussed in~\cite{CDF} are implemented. The SM background is expected to be $8663\pm332$, with the errors dominated by systematics. Since the signal does not interfere with the SM background, it is straightforward to place a bound on the underlying parameters for (\ref{proc}) if we 
naively assume the quoted errors are Gaussian. 
At the $90\%$ C.L. one finds that after all cuts the signal should predict less than $330$ events for 1 fb$^{-1}$ of data.

To proceed we first generated events for the process ~(\ref{proc}) using MadGraph/MadEvent \cite{MG} with a default generator cut of $p_T(j)>50$ GeV ~\footnote{We varied the generator-level cut to confirm the number of events passing all cuts remained the same.} and no parton-jet matching. We then passed these events to 
Pythia \cite{Pythia} with initial and final state radiation, parton showering and hadronization turned on. Hadrons were then clustered into jets with FastJet 2.4.2 \cite{fastjet}, using the anti-$k_t$ clustering algorithm with the clustering $R$ parameter set to 0.6. Events were then analyzed using Pythia, applying the CDF cuts \cite{CDF}: the missing energy in the event must be larger than 80 GeV; 
the leading jet is required to have  $p_T> 80$ GeV; a second jet must have $p_T<$30 GeV; and any additional jets must have $p_T<20$ GeV. 

The main reason for passing events to Pythia in this way was to model the production of additional jets, through ISR, FSR and hadronization, that might lead to events being rejected by the CDF cuts. The CDF analysis effectively vetoes the production of any additional hard jets, leading to a weaker bound than would otherwise be obtained without the veto.  We confirmed this effect by selecting a few points  and reapplied the analysis, alternatively switching ISR and/or FSR off. We found that switching these processes off has the effect of shifting the $p_T$ spectrum of the highest $p_T$ jet to higher values, resulting in approximately 40$\%$ to 60$\%$ more events passing the CDF cuts. Most of this effect is due to ISR. 

\begin{figure}
\begin{center}
\includegraphics[width=2.9in]{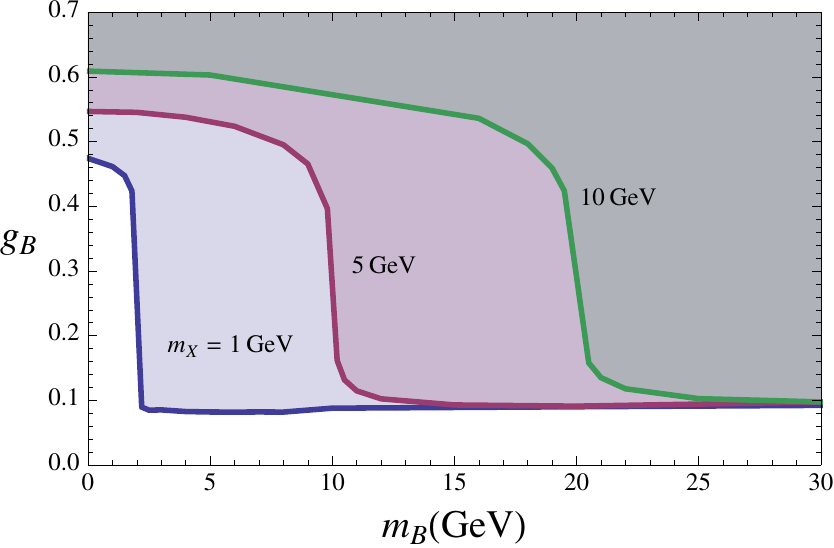}
\includegraphics[width=2.9in]{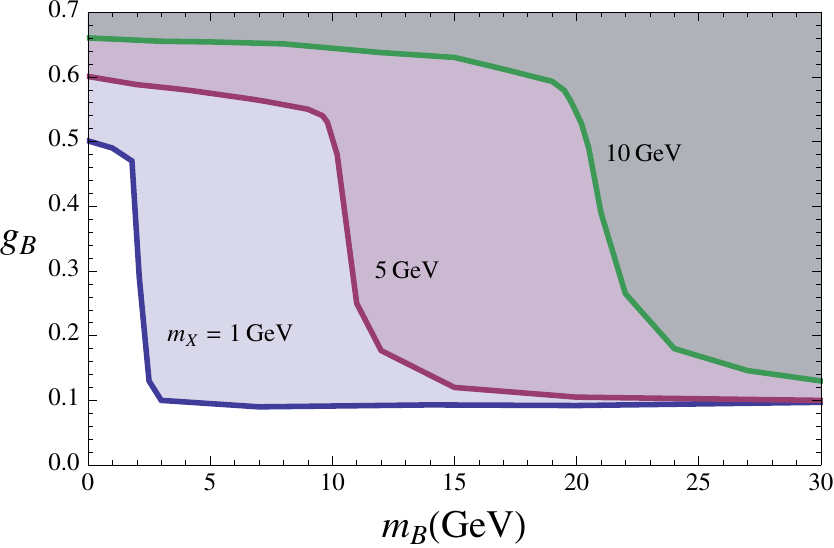}
\caption{\small 
In the left panel
we show the region in $(g_{B},m_{B})$ space that is excluded by the Tevatron for the process $p\bar{p} \rightarrow X\bar{X} + j$ for a purely vector coupling of the dark matter to the gauge boson.  The right panel shows the same exclusion but for a purely axial coupling of the dark matter to the gauge boson. 
The excluded region is shown for three different DM masses $m_{X} = 1, 5,$ and $10$ GeV, and assuming that the DM charge is $q_{V,A}=1$. Once the gauge boson $Z_{B}$ can be produced on shell the mass of the DM is irrelevant, as can be seen by the merging of three lines for $m_B\gtrsim2m_X$.  
When the DM is produced off-shell the bounds are in general weakened.  This weakening of the bounds is amplified as the DM mass increases.
\label{TeVvectoraxial}}
\end{center}
\end{figure}

In practice a parameter point $(m_B,m_X,g_B)$ was considered at the boundary of being excluded by the CDF search if between 300 and 350 signal events passed all cuts.  Accepting events in this larger range only introduces an $O(5-15 \%)$ error on the bound on the coupling, which is smaller than other errors involved. 
For a given DM mass, this  constraint translates into a exclusion region in the $(m_B,g_B)$ plane shown in Fig.~\ref{TeVvectoraxial} for either purely vector or axial coupling of the DM to the gauge boson. For definiteness we set either $q_V=1$, or $q_A=1$.  

These plots have a number of features which we now describe. First, each curve has a prominent ``transition region" that occurs at $m_B \approx 2 m_X $. Physically, this kinematic threshold divides the exclusion plots into two regions. For gauge boson masses 
$m_B < 2 m_X$
 the gauge boson is produced off-shell, and numerically we find the $Z_B$ width to be unimportant in this region. For larger masses $m_B > 2m _X$ the gauge boson is produced on-shell. In this region it is important to include the width into the amplitude consistently, since the production cross-section for the process (\ref{proc}) is highly-sensitive to it, especially to its dependence on the gauge coupling $g_B$. 
On-shell production has a larger phase space compared to the off-shell production, and the constraints in this region are correspondingly stronger, as evident by comparing the bounds at small and large masses.

For $m_B<2m_X$ the $Z_B$ boson can only be produced off-shell and  the production cross-section therefore scales as 
\ba\label{off}
\sigma_{event}=f_{\rm{off}}(m_B,m_X)g_B^4~~~~~~~\left(\rm{off-shell}\right).
\ea
In this limit our results are in agreement with~\cite{Fermi} and~\cite{Irvine}. The bounds imposed by the CDF experiment are typically less constraining than those from the invisible width of the $\Upsilon$ and the hadronic width of the $Z^0$, as well as than those from direct detection experiments.

For $m_B>2m_X$ the $Z_B$ boson can be produced on-shell and the process~(\ref{proc}) effectively becomes a two-body reaction. In this case the cross section is entirely dominated by the pole and 
\ba\label{on}
\sigma_{event}=f_{\rm{on}}(m_B)g_B^2~~~~~~~~\left(\rm{on-shell}\right).
\ea
As shown in Figure~\ref{TeVvectoraxial},  the bound on $g_B$ are now very stringent, and typically dominate over the other constraints. 
We note that the bounds shown in Fig.~\ref{TeVvectoraxial} are stronger than those reported in~\cite{Fermi}. The reason is that in~\cite{Fermi} the width of the $Z_B$ boson was assumed to be independent of $g_B$, and therefore the quadratic dependence on the coupling $g_B$ shown in~(\ref{on}) is never tested.

\begin{figure}
\begin{center}
\includegraphics[width=4.0in]{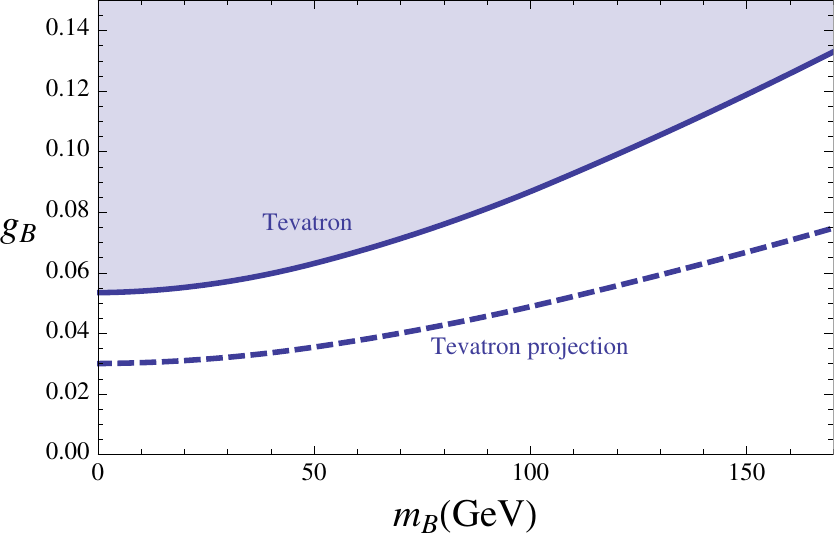}
\caption{\small Excluded area from monojet + ME events at the Tevatron assuming that the decay $Z_B\rightarrow X\overline{X}$ is allowed. Here ${\cal BR}(Z_B\rightarrow X\overline{X})=1$ for simplicity. For ${\cal BR}<1$ the bound on $g_B$ gets weakened by a factor $1/\sqrt{{\cal BR}}$. 
\label{Tev3}}
\end{center}
\end{figure}

In Figure~\ref{Tev3} we show the Tevatron bounds for on-shell $Z_{B}$ production, including the projected bounds obtained by assuming an integrated luminosity of  $\sim10$fb$^{-1}$. The bound on the signal production cross-section (after all cuts) increases by a factor $\sim\sqrt{\cal N}$, with ${\cal N} =10$.
To obtain the projections for $10$ fb$^{-1}$ of data shown in the figure we simply rescaled the current bounds of Fig.~\ref{TeVvectoraxial} by a factor ${\cal N}^{1/4}$ in the on-shell production regime $m_B>2m_X$.

Our bounds apply to generic models with $Z'$ bosons coupled to DM.  This is especially relevant given the renewed interest in $Z'$ models in light of the recently reported W$+$dijet excess by CDF~\cite{bump}.  Such a $Z'$ must couple dominantly to quarks rather than to leptons, and the connection to gauging baryon number naturally suggest itself~\cite{baryonbump}.  Arguing that the same $Z'$ may be responsible for the DM-nucleon elastic scattering cross section $10^{-40}$ cm$^{-2}$ necessary to account for the DAMA signal, the authors~\cite{baryonbump} find that the necessary gauge coupling constant $g_{B}$  is $O(0.3)$.   Rescaling the results shown in Fig~\ref{TeVvectoraxial} to account for the different branching fraction and spin of the dark matter in this model, we find a coupling of 0.3 to be close to our bound.  Future monojet + missing energy analyses using the larger Tevatron data set could exclude such coupling values.

\section{Conclusions}

In this paper we have gauged baryon number to provide a unified framework that ensures proton and dark matter stability and relates the present-day abundances of these two types of matter.

In order for the ordinary baryon number to be embedded into a gauge symmetry $U(1)_{B}$ one needs to postulate the existence of a chiral, $q'$-sector beyond the SM. A viable, perturbative formulation of this scenario also requires a dark matter particle $X$ such that processes ``$q'\rightarrow$ SM + $X$" are allowed, implying that experimental signatures of the $q'$ particles are different from a conventional 4th generation of fermions. 

The DM in these models is generally charged under the gauge baryon number $U(1)_B$, and therefore its dominant interactions with the visible sector are with the SM quarks. We discussed the collider constraints and direct detection bounds on the new leptophobic force under the assumption that the coupling of the dark matter to the new gauge boson is either purely vectorial or purely axial. We found that for a vector coupling our models are consistent with current data provided both the DM and the $U(1)_B$ force carrier have masses at the GeV scale. For purely axial couplings the direct detection bounds are weakened considerably allowing for larger dark matter masses, while the collider constraints are essentially unchanged.

A characterizing feature of our program is that the genesis of baryons and DM in the early Universe is unified.  The DM is asymmetric and its primordial asymmetry is generally comparable to the baryon asymmetry, motivating further study of GeV scale DM.

\acknowledgments
We thank Tanmoy Bhattacharya, Alex Friedland, David Morrissey, Kalliopi Petraki, and Shufang Su for discussions.  This work has been supported by the U.S. Department of Energy at Los 
Alamos National Laboratory under Contract No. DE-AC52-06NA25396. The preprint number for this manuscript is LA-UR-11-10573.

\appendix

\section{Examples of UV completions}
\label{UV}
\subsection{Model I} 

As a minimal $X$ sector we introduce a light SM singlet $X^{\pm}$ - the dark matter - in a vectorial representation of the gauge group $U(1)_B$ and 2 heavy fields $X^\pm_1$ and 
$X^\pm_2$. These have the following representations under $SU(3)_C\times SU(2)_W\times U(1)_Y\times U(1)_B$:
\ba
X^\pm\sim\left(1,1,0,\pm\left(\frac{1}{3}+\frac{1}{N}\right)\right),
\ea
and
\ba
X_1^\pm\sim\left(1,2,\pm\frac{1}{2},\pm\left(\frac{1}{3}+\frac{1}{N}\right)\right)~~~~~~~ X_2^\pm\sim\left(1,2,\mp\frac{1}{2},\pm\left(\frac{1}{3}+\frac{1}{N}\right)\right).
\ea
Then, the most general, renormalizable superpotential following from the assumed field content is $W_{tot}=W_{q}+W_{q'}+W_{\not L}+W_S+W_{X}$, with the last term given by
\ba\label{UVI}
W_X&=&m_XX^+X^-+\Lambda X_1^+X_1^-+\Lambda X_2^+X_2^-\\\no
&+&H_uX^+X_1^-+H_dX^-X_1^++Q'u_cX_1^++Qd_c'X_1^-\\\no
&+&H_uX^-X_2^++H_dX^+X_2^-+Q'd_cX_2^++Qu_c'X_2^-.
\ea
Upon integrating out the heavy fields $X^{\pm}_1$ and 
$X^{\pm}_2$ one gets Eq.~(\ref{nonrenI})
\ba\label{nonrenI'}
W_X^{\rm{eff}}&=&m_XX^+X^-+\frac{H_uQ'u_cX^+}{\Lambda}+\frac{H_dQ'd_cX^+}{\Lambda}+\frac{H_uQu_c'X^-}{\Lambda}+\frac{H_dQd_c'X^-}{\Lambda}\\\no
&+&\dots
\ea

In this specific realization, after EW symmetry breaking the DM candidate $X$ will mix with the neutral components of the fields $X^\pm_1,X^\pm_2$ and will hence acquire a coupling $O(gv_{u,d}/\Lambda)$ with the $Z^0$. This mixing is however negligible in a realistic model with $v_{u,d}/{\Lambda}\ll1$. 

Note that this field content is such that ${\rm Tr} (BY) = 0$, consistent with the assumptions of Section~\ref{Bounds}.

\subsection{Model II} 

As in the previous model here we introduce two light SM singlet chiral multiplets in a vectorial representation of the gauge group $U(1)_B$, and $SU(3)_c$ colored heavy chiral fields $Y$ and $\overline{Y}$. 
These fields have the following representations under $SU(3)_C\times SU(2)_W\times U(1)_Y\times U(1)_B$:
\ba
X\sim\left(1,1,0,\frac{2}{3}-\frac{1}{N}\right)~~~~~~~~~~\overline{X}\sim\left(1,1,0,-\frac{2}{3}+\frac{1}{N}\right),
\ea
and
\ba
Y\sim\left(\overline{3},1,\frac{1}{3},\frac{1}{3}-\frac{1}{N}\right)~~~~~~~ \overline{Y}\sim\left({3},1,-\frac{1}{3},-\frac{1}{3}+\frac{1}{N}\right).
\ea
Compared to the previous model, here the $U(1)_B$ gauge charge of the dark matter is different. 
Then the most general, renormalizable superpotential following from the assumed field content is $W_{tot}=W_{q}+W_{q'}+W_{\not L}+W_S+W_{X}$, with the last term given by
\ba\label{UVII}
W_X&=&m_X\overline{X}X+\Lambda \overline{Y}Y+u_cd'_cY+d_cX\overline{Y}.
\ea
Upon integrating out the heavy fields one gets~Eq.~(\ref{nonren})
\ba
W_X^{\rm{eff}}&=&m_X\overline{X}X+\frac{u_cd_cd_c'X}{\Lambda}+\dots
\label{nonren'}
\ea
In this model ${\rm Tr} (BY)=0$ provided new fields are added. For example one can add $Y'\sim\left(\overline{3},1,-\frac{1}{3},\frac{1}{3}-\frac{1}{N}\right)$ and $\overline{Y'}\sim\left({3},1,\frac{1}{3},-\frac{1}{3}+\frac{1}{N}\right)$.

\subsection{Chiral models} 
\label{chiralmodel}

In this section we discuss models where the dark matter candidate has a purely axial coupling to the massive gauge boson $Z_B$.  

The idea is to have two chiral fermions, say $\psi$ and $\psi^\prime$, with identical $U(1)_B$ charge. Further assume there is another 
chiral field $\phi$, with charge such that 
the Yukawa coupling 
\ba 
W= \phi \psi \psi^\prime 
\label{Wpsipsiprime}
\ea 
is gauge invariant. 
Then when $\phi$ acquires a vev, $\psi$ and $\psi'$ form a massive Dirac particle. Since $\psi$ and $\psi'$ have identical $U(1)_B$ charge, the coupling of the Dirac fermion to the massive gauge boson is purely axial \footnote{We are using 2-component notation in which all fields in the superpotential have the same chirality. In the ``left-handed" and ``right-handed" 4-component notation  $\psi_L \sim \psi$ and $\psi_R \sim \overline{\psi}^\prime$ have opposite $U(1)_B$ charge.}. 

There are two challenges to making this model more realistic. First, $U(1)_B$ gauge invariance also allows 
\ba 
\label{bad}
W=\phi \psi  \psi  + \phi \psi^\prime \psi^\prime. 
\ea 
If either of these terms are present, then after symmetry breaking $\psi$ and $\psi^\prime$ will generally form Majorana particles. If these mass terms are comparable in size to the Dirac term, then the dark asymmetry will not be preserved in the early universe. 

One way to forbid the Yukawa couplings that lead to Majorana  mass terms is impose a global or local symmetry, such that only (\ref{Wpsipsiprime}) is allowed.
 
 The other challenge is to obtain an anomaly-free theory, while keeping the dark matter chiral. This can always be achieved by the introduction of more particles. 

We now discuss a  ``proof-of-principle" model that implements these ideas. This model features a $Z_{2}$ symmetry to forbid (\ref{bad}), is anomaly-free, generates purely axial couplings for DM, and gives mass to all the fermions after the $U(1)_B$ symmetry breaking.  In general the predictions for the asymmetries are model-dependent. 

As in the models presented in the main body, a 4th generation of SM fermions is introduced to cancel the mixed $U(1)_B$-SM anomalies. Here we only discuss the particles neutral under the SM charges. The field content and charge assignments are shown  
in Table \ref{charges:chiralmodelA}.  
The cancelation of the $U(1)_B$  anomaly can be seen by noting that these fields can be arranged as 
\ba 
{\bf 1}_{+4}+  {\bf 4}_{-1} +  {\bf 6}_{+2} + {\bf \overline{4}}_{-3}
\ea 
which is the decomposition of the (anomaly-free) ${\bf 5} + {\bf \overline{10}}$ of $SU(5)$ under its $SU(4) \times U(1)$ subgroup. The only vector particles are 
$Z_{-4}$ and $\overline{Z}_4$ introduced to allow Yukawa couplings for the charge 2 chiral fields. There is also a single $Y$ introduced to assist in giving mass to all the fermions. 

\begin{table}
\begin{center}
\begin{tabular}{|c|c|c||}
\hline
 &  $U(1)_B$ & $Z_2$   \\ \hline 
$\Phi_4$ &$+4$ & $1$   \\ \hline 
$\Phi_{(-1;i)}$ & $-1$ &  $1$   \\ \hline 
$\psi_{(2;k)}$ & $2$ &  $-1$   \\ \hline 
$\psi^\prime_{(2;k)}$ & $2$ & $+1$  \\ \hline 
$\Phi_{(-3;j)}$ & $-3 $ & $1$ \\ \hline \hline
$Z_{-4}$  & $-4$ & $-1$ \\ \hline 
$\overline{Z}_{4}$  & $4$ & $1$ \\ \hline 
$Y$ & $0$ & $-1$ \\ \hline 
\end{tabular}
\caption{Model A: gauge $U(1)_B$ and exotic  $Z_2$ charges of the dark matter sector. The subscripts on the fields indicate their $U(1)_B$ charges. $i,j=1,..,4$ and $k=1,..,3$. All particles are neutral under the SM. \label{charges:chiralmodelA}}
\end{center}
\end{table}

We next make a few comments about  scales. We assume that SUSY breaking occurs first in the visible sector, and then is communicated to the dark matter sector, possibly by $U(1)_B$ gauge interactions and/or higher dimension operators (such as the transfer operators), leading to
$U(1)_B$ symmetry breaking and SUSY mass splittings in these fields at the GeV to tens of GeV scale. We also assume that the mass scales appearing in the dark matter  superpotential are of this size.


In this model we decompose the ${\bf 6}_{+2} $ into three generations of $\psi_{(2;k)}$ and $\psi^\prime_{(2;k)}$, where $\psi$ and $\psi^\prime$ have opposite $Z_2$ charges.   These fields play the role of the $\psi$ and $\psi^\prime$ in the toy model. The role of $\phi$ is played by $Z_{-4}$. To cancel  anomalies we also introduce a vector-partner, $\overline{Z}_4$.
After $U(1)_B$ symmetry breaking the $\psi$ and $\psi^\prime$ fields will marry to become a Dirac particle. 

The most general renormalizable superpotential allowed by these symmetries is 
\ba 
W &=& \lambda_{ij} (\Phi_4 + \overline{Z}_4)  \Phi_{(-3;i)} \Phi_{(-1;j)} + \lambda^{\prime}_{ij k}  \psi^\prime_{(2;k)} \Phi_{(-1;i)} \Phi_{(-1;j)} \nonumber \\
 & & + \lambda^{\prime \prime \prime}_{kl} Z_{-4}  \psi_{(2;k)} \psi^\prime _{(2;l)}   + Y Z_{-4} (\Phi_4 +\overline{Z}_4)
 + \mu Y^2  
 \label{WchiralmodelA}
\ea 
where $k,l=1,..,3$ and $i,j=1,...,4$.  

We assume that SUSY breaking leads to $U(1)_B$ breaking vevs for $Z_{-4},\overline{Z}_4, \Phi_4$ and $Y$, but not any of the other fields (otherwise $\psi$ and $\psi^\prime$ will not in general form Dirac particles). With this assumption all of the fermions become massive: $\psi$ and $\psi^\prime$ form 3 generations of Dirac particles, with purely axial and universal couplings to $Z_B$; the $\Phi_{(-3;i)}$ and $\Phi_{(-1;j)}$ mix to form 4 Dirac particles with both universal vector and axial couplings to $Z_B$; and $Y$, $Z_{-4}$ and $\overline{Z}_4$ mix to form 3 Majorana particles, with non-universal couplings to $Z_B$. We will refer to the massive Dirac eigenstates as $\psi_a$, $\psi^\prime_b$, $a,b=1,2,3$, and $\Phi_{(-3;m)}$, $\Phi_{(-1;m)}$, $m=1,...,4$, or $\psi_a$ and  $\Phi_m$ for short. 

These vevs leave four accidental symmetries. 

There are three  generalized $\psi$ particle numbers, one for each $\psi_a$,  in which $\psi_a$ has charge $+1$ and $\psi^\prime_a$ has charge $-1$. In addition,  the $\Phi_{(-3;m)}$'s and  $\Phi_{(-1;m)}$'s have opposite charge equal to half of the $\psi_a$. The lightest particle of each generalized $\psi_a$ number will be stable.

There is also an unbroken ${\cal R}-$parity for which $\psi_{(2;i)}$ and $\psi^\prime_{(2;i)}$ are even, and $\Phi_{(-3;i)}$ and $\Phi_{(-1;i)}$ are odd. The lightest ${\cal R}_p$ odd particle is stable. If we make the simplifying assumption that all the scalars are heavier than the fermions, then the lightest ${\cal R}_p$  odd particle will either be the $U(1)_B$ gaugino or the lightest (fermionic) $\Phi_m$. 

This model therefore has from one to five dark matter candidates, depending on the relative masses of the $U(1)_B$ gaugino, the $\psi_a$'s, and the lightest $\Phi_m$.  The lightest $\Phi_m$ is always stable since it has $\psi_a$ number $1/2$ and there are no other lighter particles with fractional $\psi_a$ number for it to decay into.  From the conservation of the individual $\psi_a$, the lightest $\psi_a$ can in principle only decay to 2 $\Phi_m$'s and a third $\psi_a$-neutral fermion (needed to conserve spin). A sufficient condition for the lightest $\psi_a$ to be stable if it is lighter than twice the mass of the lightest $\Phi_m$.  Depending on the kinematics, these circumstances can add up to 3 more stable particles. Finally, if the gaugino is the lightest ${\cal R}_p$ odd particle then it too is stable. 

Since the dark matter sector is more involved, the role of $X$ in the transfer operators (\ref{nonrenI}) or (\ref{nonren}) can in principle be filled by any of the fields in the dark sector, subject only to the requirement that the transfer operator is $U(1)_B$ and $Z_2$ invariant. 

Up to this point we have not discussed the relative normalization between the $U(1)_B$ charges of the dark matter sector and those of the visible and exotic quarks. 
In fact, this normalization is model-independent.  
This is because the dark matter sector alone is automatically $U(1)_B$ anomaly-free. We are therefore free to choose the normalization appropriately such that the transfer operators (\ref{nonrenI}) or (\ref{nonren}) are $U(1)_B$ gauge invariant. 

Next, the transfer operators should be $Z_2$ invariant. To do that we extend the $Z_2$ to act on the exotic quarks  such that they are odd under the $Z_2$: $Q^\prime \rightarrow - Q^\prime$, $u^{c \prime} \rightarrow - u^{c \prime}$, 
$d^{c  \prime} \rightarrow -  d^{c \prime}$. The exotic leptons and SM particles, including the Higgses, are assumed to be even.  

Under these conditions only $\psi_{(2;k)}$ or $Z_{-4}$ have the appropriate quantum numbers to couple linearly to the visible and exotic quarks.  
Then with an appropriate universal rescaling of the dark matter sector $U(1)_B$ charges, transfer operators of the form 
(\ref{nonrenI}) or (\ref{nonren}) are allowed with $X= \psi_{(2;k)}$ or, with a different rescaling, $X=Z_{-4}$. 

We now focus on the scenario in which the lightest $\psi_a$ is light enough to be stable. 
Then this particle is a  dark matter candidate with a purely axial coupling to $Z_B$. The predictions 
for the dark matter asymmetries are dependent however on the rest of the spectrum, and on the identity of the particle $X$ appearing in the transfer operator. 

There is one circumstance however, in which our chemical potential analysis appearing in Section \ref{chempot} carries through exactly. Namely, if in the transfer operator $X= \psi_{(2;k)}$ and if in addition 
 there is a modest mass gap between the lightest $\psi_a$ and the rest of the dark matter spectrum. Then the present-day abundances and asymmetries of the heavier stable particles will be Boltzmann suppressed compared to the lightest $\psi_a$.

\section{Chemical potential analysis}
\label{chempot}

Let us focus on the case where the AD condensate evaporates before the sphalerons decouple, namely  $T_\phi>T^*$.

As already emphasized, the precise value of the present-day asymmetries in this case is set by thermodynamical relations at the sphaleron scale $T^*$. Once the ratio $\eta_{B_q}/\eta_{B_{q'}}$ is calculated at the sphaleron scale $T^*$ one can obtain the present-day asymmetries for the DM and the ordinary SM baryons using eq.~(\ref{note1}).  

Since we expect $T^*=O(100$ GeV$)$, the $N$ exotic generations (with $U(1)_B$ charge $-1/N$) and the top are assumed \emph{not} decoupled in the following analysis. For simplicity, we will only consider the contribution of the chiral fields, as the vectorial representations, as well as all SUSY partners, are all taken to be heavy enough so that their contribution to the following chemical potential analysis is negligible. An exception is the DM particle $X$ which in these models must have a mass below $O(100$ GeV$)$ if it is to have the correct present-day cosmological abundance.

As it is not known whether the sphaleron scale $T^*$ is above or below the EW phase transition scale $T_{EW}$, we will analyze both cases $T^*<T_{EW}$ and $T^*\geq T_{EW}$ in turn. The former case is expected if the EW transition is smooth, whereas the latter if the transition is strongly first order. We will see that the ratio between the dark matter $\eta_X$ and baryon $\eta_{B_q}$ asymmetries \emph{does not} depend on the precise relation between the scales $T^*,T_{EW}$ if the $q'$ masses are in the few hundred GeV range. Under this latter hypothesis we thus expect the ratio~(\ref{note1}) to be the same in models with a first order or second order EW phase transition.

In the following subsection we will calculate the ratio of $B_{vis}$ asymmetry to $X$ asymmetry for model I ~(\ref{nonrenI}) and model II (\ref{nonren}). Our analysis can be generalized to arbitrary $U(1)_D$-invariant interactions mediating $q'$ decay. 

\subsection{Low-temperature sphaleron decoupling}

We first assume that the sphalerons decouple below the EW phase transition. We closely follow the formalism of~\cite{HT} and set the chemical potential of the physical, CP-even Higgs to zero
\ba\label{mu0}
\mu_0=0~~~~~~~~~~~\left({\rm{for}}~~~T^*<T_{EW}\right).
\ea 
It then follows that the chemical potentials of the left and right fermions of the same flavor equal to each other. Similarly, we equate the chemical potentials of the transverse and longitudinal components of the $W^\pm$ boson to $\mu_W$. As long as the transfer operators of model I~(\ref{nonrenI'}) or model II (\ref{nonren'}) are in chemical equilibrium we have 
\ba\label{analog}
\mu_{d'}&=&\mu_d-\mu_X~,~~~~~~~~~~~~~~~~~~~~\mu_{d'}=\mu_W+\mu_{u'}~,~~~~~~~~~~~\mu_{e'}=\mu_e,~~\hbox{(model I)} \no \\
\mu_{d'}&=&-\mu_u-\mu_d+\mu_X~,~~~~~~~~~~~\mu_{d'}=\mu_W+\mu_{u'}~,~~~~~~~~~~~\mu_{e'}=\mu_e,~~\hbox{(model II)}
\ea
where $\mu_{u,d,u',d'X}$ are the chemical potentials of the up- down-type quarks and exotic quarks, as well as of the DM, while $\mu_{e,e'}$ those of the charged leptons. $R$-parity, lepton-flavor violating effects are assumed to be in equilibrium so that the chemical potentials of the various generations coincide and so that those of the neutrinos vanish.

We aim to solve a system of 6 unknown $\mu_u, \mu_d,\mu_\nu,\mu_e,\mu_W,\mu_X$ with 5 constraints. The first constraint comes from the vanishing of the total electric charge in the Universe; the second and third constraints relate the up and down elements of the weak doublets as well as $\mu_W$ and comes from the $W^\pm$ exchange; the fourth involves left-handed fermions only and comes from the sphaleron process associated with the $B_{q}$ and $B_{q'}$ anomalies; and finally the fifth constraint reads $\mu_\nu=0$ and follows from the assumption that lepton violating operators responsible for the generation of Majorana masses for the SM neutrinos are in chemical equilibrium. 

Working at leading order in the chemical potentials as in~\cite{HT}, the 5 constraints respectively read:
\ba\label{chem}
\eta_Q&\propto&3\cdot3\cdot\frac{2}{3}(2\mu_u)-3\cdot3\cdot\frac{1}{3}(2\mu_d)-3(2\mu_e)-3\cdot2\mu_W\\\no
&+&\left[N\cdot3\cdot\frac{2}{3}(2\mu_{u'})-N\cdot3\cdot\frac{1}{3}(2\mu_{d'})-N(2\mu_{e'})\right]f(M_{q'}/T^*)\\\no
&\equiv&0\\\no
\mu_W=\mu_d-\mu_u&=&\mu_e-\mu_\nu\\\no
3[\mu_u+2\mu_d+\mu_\nu]&=&-N[\mu_{u'}+2\mu_{d'}+\mu_{\nu'}]\\\no
\mu_\nu=\mu_{\nu'}&=&0.
\ea

In the above expression the function $f(M_{q^\prime}/T^*)$ accounts for the mass dependence of the $q^\prime$ contribution, which is taken for simplicity to be the same between the up-type and down-type (which is preferred by the electroweak precision constraints) and for any (exotic) flavor. This function is given by
\ba
f(x)=\frac{3}{2\pi^2}\int_0^\infty dy\,\frac{y^2}{\cosh^2\left(\frac{1}{2}\sqrt{x^2+y^2}\right)}\leq1,
\ea
and it is normalized so that $f(0)=1$. At temperatures above the electroweak phase transition $x=0$ since the exotic quarks obtain all of their mass from electroweak symmetry breaking. 

Note that in the charge asymmetries bosons count twice as much as fermions~\cite{HT}, but fermions can have both left and right components (see round brackets in the above expressions).

After the AD condensate has evaporated  
the primordial asymmetry for $U(1)_D$ is completely carried by the fundamental particles and is given by \footnote{$\eta_D$ is intimately connected with the asymmetry $\eta_S$ carried by the Higgs fields $S,\overline{S}$. Indeed, from~(\ref{gaugeB} and \ref{D}) one has
\ba\label{charges}
\eta_B=\eta_D+B(S)\eta_{S}.
\ea
If the $U(1)_B$ gauge symmetry is unbroken, then $\eta_B=0$ and consequently $\eta_D=-B(S)\eta_{S}\neq0$. However, after $U(1)_B$ symmetry breaking the reactions $s^0 \longleftrightarrow Z_B Z_B$ quickly lead to 
$\mu_S=0$ and no asymmetry is carried by the $S \equiv s^0 + i s^I$ field. Then the $s^0$ field decays to jets. 
The baryon Higgs sector therefore does not contribute to either of the $\eta_{B_{q,q^\prime}}$ asymmetries.}
\ba
\eta_D&\propto&3\cdot3\cdot\frac{1}{3}(2\mu_u)+3\cdot3\cdot\frac{1}{3}(2\mu_d)\\\no
&-&\left[N\cdot3\cdot\frac{1}{N}(2\mu_{u'})+N\cdot3\cdot\frac{1}{N}(2\mu_{d'})\right]f_{q^\prime}+B(X)(2\mu_X f_X)  \\\no
&\neq&0,
\ea
where  for brevity $f_{q^\prime} \equiv f(M_{q^\prime}/T^*)$ and $f_X \equiv f(m_X/T^*)$. From Table \ref{table:globalcharges}, $B(X)=1/3+1/N$ (model I) or $B(X)=2/3-1/N$ (model II).
One can show that the non-vanishing of $\eta_D$ requires  the chemical potentials $\mu_{q,q^\prime,X}$ to all be nonzero. 

Evaluating the  $U(1)_{B_{q,q^\prime}}$ asymmetries at $T=T^*$ gives 
\ba
\label{chargesQ}
\eta_{B_q}&\propto&3\cdot3\cdot\frac{1}{3}(2\mu_u)+3\cdot3\cdot\frac{1}{3}(2\mu_d)+B_q(X)(2\mu_X f_X),\\ 
\eta_{B_{q'}}&\propto&\left[-N\cdot3\cdot\frac{1}{N}(2\mu_{u'})-N\cdot3\cdot\frac{1}{N}(2\mu_{d'})\right]f_{q^\prime}+B_{q^\prime}(X)(2\mu_X f_X) 
\label{chargesQp}
\ea
Next we insert into~(\ref{chargesQ}) and (\ref{chargesQp}) the explicit solution of~(\ref{chem}) (making use of~(\ref{analog})), to obtain an expression for 
$\eta_{B_q}/\eta_{B_{q^\prime}}$. Then 
since these asymmetries are conserved below the sphaleron scale $T^*$, we evaluate them at the present-day era using (\ref{present-Bq}) and (\ref{present-Bqp}).
We then find the present-day ratio~(\ref{note1}) to be 
\ba\label{note3f}
\frac{\eta_{B_{vis}}}{\eta_X}&=&-\frac{9 (-1 + f_{q^\prime}) N (21 + 5 f_{q^\prime}N))}{ 99 + (33 + 618 f_{q^\prime}) N + f_{q^\prime} (17 + 135 f_{q^\prime}) N^2)}
~~\hbox{(model I)}
\ea
\ba\label{note2f}
\frac{\eta_{B_{vis}}}{\eta_X}&=&-\frac{9 N(21+f_{q^\prime}(45+2(3+5f_{q^\prime})N)}{99+N(-66+f_{q^\prime}(618+(-7+135f_{q^\prime})N))}  ~~~~\hbox{(model II)} 
\ea
where we have set $f_X=1$.

\subsection{High-temperature sphaleron decoupling}
\label{HTSD}

In the case the sphalerons decouple above or at the EW phase transition we again follow~\cite{HT} and impose, instead of~(\ref{mu0}), the vanishing of the chemical potential of the $W^\pm$:
\ba\label{above}
\mu_W&=&0~~~~~~~~~~~~~~~~~~~~~\left({\rm{for}}~~~T^*\geq T_{EW}\right)\\\no
&=&\mu_{dL}-\mu_{uL}\\\no
&=&\mu_{\nu L}-\mu_{eL}.
\ea
Similar relations hold for the $q^\prime$ fields. 

Now there is no distinction between the up and down elements of the weak doublet, but we should make a distinction between left and right components of the chiral fields, as well as between the ``up-type" $(\mu_{H_u})$ and ``down-type" $(\mu_{H_d})$ Higgs bosons. The $B-$term however enforces $\mu_{H_u} = -\mu_{H_d} \equiv \mu_0$, so the two Higgs boson chemical potentials are opposite. 

The Yukawa couplings then imply
\ba\label{Higgs}
\mu_0=\mu_{uR}-\mu_{uL}&=&\mu_{dL}-\mu_{dR}=\mu_{eL}-\mu_{eR},
\ea
and similarly for the $q^\prime$ fields.  Yukawa couplings between the 4th and first three generations leptons  imply
\ba
\label{Higgsprime}
\mu_0=\mu_{eL}-\mu_{e^\prime R}= \mu_{e^\prime L}-\mu_{e R},
\ea
which when combined with the previous Yukawa interactions implies 
\ba \label{leptons1}
&&\mu_{e^\prime L,e^\prime R}=\mu_{eL,eR}\no
\ea
In addition, the interactions $ H_u L^\prime \nu^\prime_c$ and $S \overline{S} \nu^{\prime }_c $ (or $\nu^{\prime }_c \nu^{\prime }_c$) give
\ba\label{leptons2}
&&\mu_{\nu^\prime R}=\mu_0+\mu_{\nu^\prime L} \\
&& \mu_{\nu^\prime R}=0 
\ea

As before, the relevant constraints come from electric change neutrality, the sphaleron process, and the $R_p$ lepton violating processes $LQD^c$ and $LLE^c$:
\ba\label{chem'}
\eta_Q&\propto&3\cdot3\cdot\frac{2}{3}(\mu_{uL}+\mu_{uR})-3\cdot3\cdot\frac{1}{3}(\mu_{dL}+\mu_{dR})\\\no
&-&3(\mu_{eL}+\mu_{eR})+2\mu_0\\\no
&+&N\cdot3\cdot\frac{2}{3}(\mu_{u^\prime L}+\mu_{u^\prime R})-N\cdot3\cdot\frac{1}{3}(\mu_{d^\prime L}+\mu_{d^\prime R})\\\no
&-&N(\mu_{e^\prime L}+\mu_{e^\prime R}) \\\no
&\equiv&0\\\no
3[3\mu_{uL}+\mu_{\nu L}]&=&-N[3\mu_{u^\prime L}+\mu_{\nu^\prime L}]\\\no
\mu_{\nu L}+\mu_0&=&0 
\ea  
plus the condition~(\ref{above}), Higgs exchange eqs.~(\ref{Higgs}) and (\ref{Higgsprime}), and the relations~(\ref{leptons1}) and (\ref{leptons2}). Note that here $f=1$ for all the exotic 
fermions since they only acquire a mass from electroweak symmetry breaking.  
The last equation in (\ref{chem'}) follows from the equilibrium conditions of the $R_p$ lepton-violating operators.

The explicit solution to these equations is:
 \ba\label{muWE}
\mu_u=\mu_d,~~~~~\mu_{u'}=\mu_{d'}=-\frac{3}{N}\mu_u,~~~~~\mu_\nu=\mu_e=\mu_{\nu'}=\mu_{e'}=0, ~~~\mu_{0}=0.
\ea
This implies that left and right components have the same chemical potential.
As before, we can evaluate the asymmetries at $T=T^*$ and then match at temperatures below the decoupling temperature of the transfer operator. 
 It then follows that the ratio between the primordial asymmetries is for model I, 
\ba \label{note2a}
\frac{\eta_{B_{vis}}}{\eta_X}&=&\left[\frac{12+2(1+\frac{3}{N})B_q(X)}{12 (\frac{3}{N})+ 2(1+\frac{3}{N})B_{q^\prime}(X)}\right] B_{q^\prime}(X)- B_q(X) =0
\nonumber
\ea 
since for the transfer operator of model I, $B_{q^\prime}(X)=\frac{3}{N} B_q(X)$. 
While for model II, 
\ba\label{note2}
\frac{\eta_{B_{vis}}}{\eta_X} &=&  \left[\frac{12+2(2-\frac{3}{N})B_q(X)}{12 (\frac{3}{N})+ 2(2-\frac{3}{N})B_{q^\prime}(X)}\right] B_{q^\prime}(X)- B_q(X)   \approx -1. \no
\ea
The last step is a good approximation for any $N$. 

Interestingly this same solution~(\ref{muWE}) is obtained from the {\it low} temperature sphaleron decoupling assumption with $f_{q^\prime}=1$.  In other words, if the sphalerons decouple {\it above} the electroweak phase transition, i.e., $T^* > T_{EW}$, the present-day asymmetries are obtained from eqs.~(\ref{note3f}) and (\ref{note2f}) by setting $f_{q^\prime}=1$.

The result~(\ref{muWE}) states that the lepton number is washed out by the equilibrium processes, while the baryon number \emph{is not} because it is embedded into the nonanomalous symmetry $U(1)_D$. The washing out of the lepton number follows as the combined effect of perturbative \emph{and} nonperturbative lepton violation, the latter processes due to the sphalerons. If one of these two ingredients are not in play, the chemical potentials of the leptons are nontrivial in general.

\bibliography{admbib}
\bibliographystyle{JHEP}

 \end{document}